\documentclass[preprint,superscriptaddress,amssymb,pra,aps]{revtex4-2}
\usepackage{graphicx,xcolor,transparent,mathtools,amsmath,amsthm,soul,changes}
\newcommand{\la}{\langle}
\newcommand{\ra}{\rangle}

\begin{document}
\title{Increased success probability in Hardy's nonlocality: Theory and demonstration}

\author{Duc Minh Tran\footnote{Current address: Laboratoire ICB, UMR CNRS, Universite de Bourgogne Franche-Comte, Dijon, France}}
\affiliation{Nano and Energy Center, University of Science, Vietnam National University, Hanoi, 120401, Vietnam}
\author{Van-Duy Nguyen}
\affiliation{Phenikaa Institute for Advanced Study, Phenikaa University, Hanoi 12116, Vietnam}
\author{Le Bin Ho}
\affiliation{Frontier Research Institute for Interdisciplinary Sciences, Tohoku University, Sendai 980-8578, Japan}
\affiliation{Department of Applied Physics, Graduate School of Engineering, Tohoku University, Sendai 980-8579, Japan}
\author{Hung Q. Nguyen}
\email{hungngq@hus.edu.vn}
\affiliation{Nano and Energy Center, University of Science, Vietnam National University, Hanoi, 120401, Vietnam}
\begin{abstract}
Depending on the way one measures, quantum nonlocality might manifest more visibly. Using basis transformations and interactions on a particle pair, Hardy logically argued that any local hidden variable theory leads to a paradox. Extended from the original work, we introduce a quantum nonlocal scheme for $n$-particle systems using two distinct approaches. First, a theoretical model is derived with analytical results for Hardy's nonlocality conditions and probability. Second, a quantum simulation using quantum circuits is constructed that matches very well to the analytical theory. When demonstrated on real quantum computers for $n=3$, we obtain reasonable results compared to theory. Even at macroscopic scales as $n$ grows, the success probability asymptotes 15.6\%, which is stronger than previous results.
\end{abstract}
\maketitle
	
\section{Introduction}
Through a gedanken experiment on particle-antiparticle interactions inside two intertwined Mach–Zehnder interferometers, Hardy delivered a proof of Bell-nonlocality without using inequalities \cite{hardy1992,hardy1993}, an all-versus-nothing criterion for \emph{local hidden variable} (LHV) theory. In his setting, a particle and its antiparticle, most commonly an electron and a positron, are sent through two Mach-Zehnder interferometers that cross each other at one of their paths, and will annihilate should they meet. This arrangement is represented by Eq.~\eqref{ori1} and \eqref{ori2}. However, if there is no positron, the electron would always go to its constructive interference detector, and vice versa, per Eq.~\eqref{ori3} and \eqref{ori4}. Hardy originally argued that there is a discrepancy between LHV models and quantum predictions. Due to the nonlocal nature of quantum states, it is possible to argue the pair met without an annihilation event, causing a paradox, as in Eq.~\eqref{ori5}. In details, suppose we have two physical observables $ U_i = |u_i\ra\la u_i|$ and $ D_i = |d_i\ra\la d_i|$ with their basis transformation
\begin{equation*}
	\begin{aligned}
		|u_{i}\ra&=A^{*}|c_{i}\ra-B|d_{i}\ra, \ \  |v_{i}\ra=B^{*}|c_{i}\ra+A|d_{i}\ra,\\
		|c_k\ra&=A|u_k\ra+B|v_k\ra, \ \ |d_k\ra=-B^*|u_k\ra+A^*|v_k\ra,
	\end{aligned}
\end{equation*}
where $|A|^2 + |B|^2 = 1$. With $N$ being the normalization factor, the quantum state writes in equivalent forms:
\begin{align}
	|\Psi\ra&=N(|c_{1}\ra|c_{2}\ra-A^{2}|u_{1}\ra|u_{2}\ra) \label{ori1} \\
	&=N(A B|u_{1}\ra|v_{2}\ra+A B|v_{1}\ra|u_{2}\ra+B^{2}|v_{1}\ra|v_{2}\ra) \label{ori2} \\
	&=N(|c_{1}\ra(A|u_{2}\ra+B|v_{2}\ra)-A^{2}(A^{*}|c_{1}\ra-B|d_{1}\ra)|u_{2}\ra) \label{ori3} \\
	&=N((A|u_{1}\ra+B|v_{1}\ra)|c_{2}\ra-A^{2}|u_{1}\ra(A^{*}|c_{2}\ra-B|d_{2}\ra)) \label{ori4} \\
	&=N(|c_{1}\ra|c_{2}\ra-A^{2}(A^{*}|c_{1}\ra-B|d_{1}\ra)(A^{*}|c_{2}\ra-B|d_{2}\ra)). \label{ori5}
\end{align}
We split these arguments into three sets of Hardy's nonlocality conditions, following the nomenclature in \cite{kar1997_nonlocality_term, ghosh1998, Cabello2013}. Firstly in Eq.~\eqref{ori1} and \eqref{ori2}, the probability to find $U_1 = 1$ and $U_2 = 1$ simultaneously is zero: $P(U_1 U_2) = 0$. Here and throughout the paper, we write $P(U_1 U_2 \cdots) = 0$ as a shorthand for $P(U_1 U_2 \cdots = 1) = 0$. Secondly in Eq.~\eqref{ori3} and \eqref{ori4}, LHV-correlations between $ U_i$ and $ D_i$ are exposed by mixing measurement basis for different particles. Using conventional notations for conditional probability, the probability that $U_2 = 1$ under the condition of $D_1 = 1$ is $P(U_2|D_1) = 1$, due to the absence of the $|d_1\ra |v_2\ra$ term in Eq.~\eqref{ori3}. Similarly, $P(U_1| D_2) = 1$. Finally in Eq.~\eqref{ori5}, $P( D_1 D_2) > 0$, when combined with the secondly established probabilities would imply $P(U_1 U_2) > 0$. This last nonzero probability directly contradicts the first condition, which is $P(U_1 U_2) = 0$.
Hence, $P(D_1D_2)$ is the probability of success for demonstrating nonlocality in Hardy's context and we refer to it as the ``success probability" $P_\text{success}$.

This elegant approach was systematically extended to many aspects of 2-particle states \cite{Popescu1992, cabello1999, Zukowski2002, YangPRA19,VallonePRA11,JordanPRA94}, especially without the maximally entangled state \cite{goldstein1994}. Generalizations for 3-particle states \cite{wu1996,wu2000}, and arbitrary $n$-particle states \cite{kar1997, ghosh2010, pagonis1992, cereceda2004, jiang2018} was demonstrated, with the maximum probability of the nonlocal state on a general 3-particle state reached 12.5\% \cite{ghosh1998}. The Hardy-type nonlocality can be proved for GHZ states \cite{wu2000, cereceda2004, jiang2018}, graph states \cite{Cabello2008,Gachechiladze2015}, symmetric states \cite{Wang2012}, $W$ and Dicke states \cite{Barnea2015}, and Wigner's argument \cite{Home2015}. Its strength and visibility compared to other nonlocality proofs such as Bell's theorem, GHZ and CHSH are explored \cite{Garuccio1995,vanDam2005,Ghirardi2008,  Braun2008}, and there are several unification attempts \cite{Mancinska2014, Dong2020}. The nonlocal visibility can be amplified using ``ladder" logic \cite{boschi1997, Barbieri2005,Cabello2013}, graph-theoretic logic \cite{Sohbi2019,Svozil2021}, and high-dimensional systems \cite{chen2013, Chen2017, Meng2018,Rabelo2012}. Meanwhile, the correlation between nonlocality and entanglement is at the spotlight in quantum foundation researches \cite{goldstein1994, Acin2005, Brunner2005, Junge2011, Liang2011, Vidick2011, Diley2018}.  More importantly, there are experimental evidences in photons \cite{irvine2005,Lundeen2009,Yokota2009,luo2018}, atoms \cite{matsukevich2008,hofmann2012}, and generic quantum computers \cite{das2020,Hou2021}.

To derive a Hardy's type paradox, one starts with choosing a specific quantum state, such as GHZ or W. After establishing a set of Hardy's nonlocality conditions for the chosen state, the success probability is calculated verifying these assumptions. In this work, we follow this strategy and find a stronger version of n-particle Hardy's paradox on the measure of the success probability. It approaches 15.6\% as the system size $n$ grows, compared to previous works with vanishing success probability at high $n$ limit \cite{cereceda2004, jiang2018}. Besides the theoretical description, we provide a quantum simulation by using quantum circuits on any simulators and any available quantum hardware. We execute this simulation on IBM quantum computers and obtain data that match reasonable well to the theory. 

\section{$n$-particle Hardy's paradox}

\begin{figure}[t]
	\begin{center}
		\includegraphics[width=0.8\textwidth]{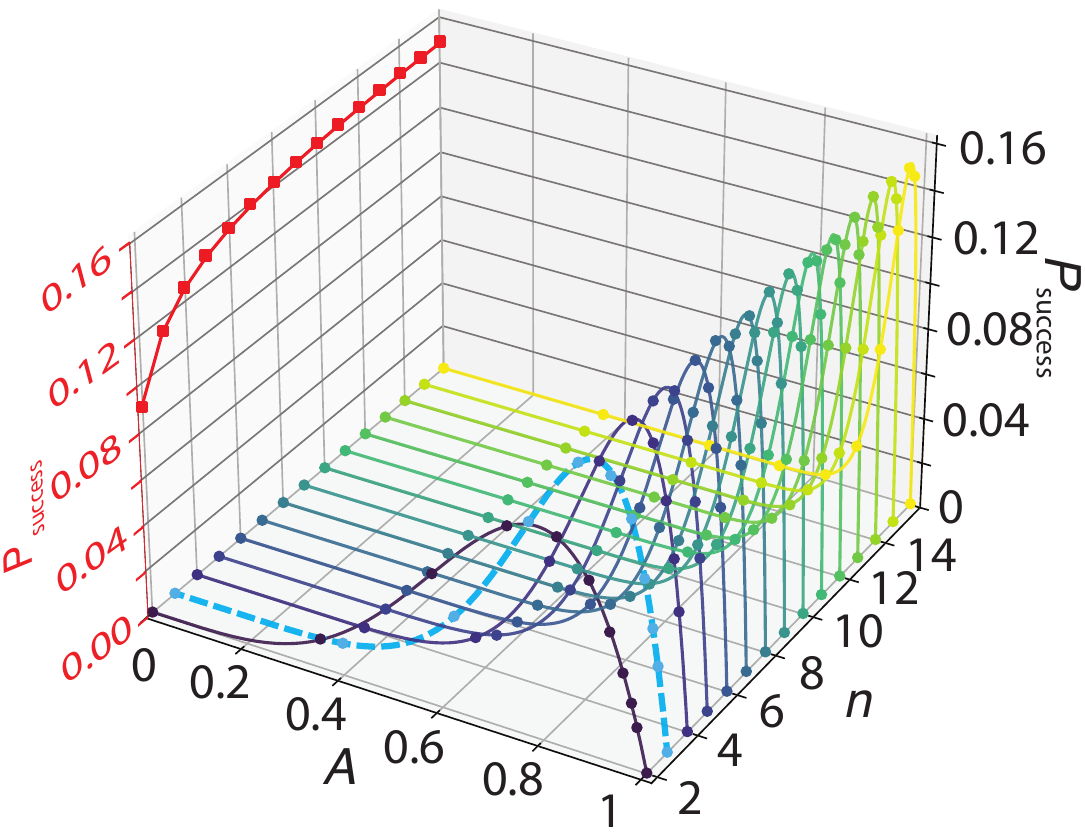}
		\caption{\textbf{Analytical result:} Success probability $P_\text{success}$ as a function of the transformation coefficient $A$ and different system size $n$. The red curve with square dots on the left is a projection of the maximum $P_\text{success}$ for different $n$. The cyan dashed curve highlights an example $n=3$. Solid lines are analytical result Eq.~\eqref{eq:Gammasame}, while dots are simulation result obtained from the QASM quantum simulator provided by IBM.}
		\label{fig:fig1psi}
	\end{center}
\end{figure}
	
Let us consider a system of $n$ qubits labeled by $\Omega = \{1,\cdots,n\}$. For each qubit $k \in \Omega$, we introduce two non-commuting pairs of observables $ U_k = |u_k\ra\la u_k|$, $ V_k = |v_k\ra\la v_k|$ and $ C_k = |c_k\ra\la c_k|$, $ D_k = |d_k\ra\la d_k|$. These observables span in two orthogonal bases, $\{|u_k\ra, |v_k\ra\}$ and $\{|c_k\ra, |d_k\ra\}$, such that $\la u_k|v_k\ra = \la c_k|d_k\ra = 0$, and obey
\begin{align}
	|u_{k}\ra &=A_k^*|c_k\ra-B_k|d_k\ra,\ |v_k\ra=B_k^*|c_k\ra+A_k|d_k\ra, \label{eq:bs1}\\
	|c_k\ra&=A_k|u_k\ra+B_k|v_k\ra ,\	|d_k\ra=-B_k^*|u_k\ra+A_k^*|v_k\ra, \label{eq:bs2}
\end{align}
where $A_k, B_k$ are complex coefficients satisfying $|A_k|^2 + |B_k|^2 = 1$. Following notations in Ref.~\cite{jiang2018}, we denote ${\mathcal{U}}_\alpha \equiv \Pi_{k\in \alpha} U_k$ and ${\mathcal{D}}_\alpha \equiv \Pi_{k\in \alpha} D_k$ for any subset $\alpha \subseteq \Omega$, and likewise $\mathcal{A}_\alpha \equiv \Pi_{k\in\alpha}A_k$ and $\mathcal{B}_\alpha \equiv \Pi_{k\in\alpha}B_k$. In our indices, Latin letters denote numbers and Greek letters denote sets of number. For an arbitrary set $\alpha$, let $|\alpha|$ be its cardinality - i.e. the number of elements in $\alpha$, $\mathcal{P}(\alpha)$ be its power set, and $\overline{\alpha} = \Omega \backslash \alpha$.

Our state of interest can be viewed as a general $n$-particle state in the $\{|u\ra_k,|v\ra_k\}^{\otimes n}$ basis without the $|u_1\ra \otimes |u_2 \ra \otimes \cdots \otimes |u_n\ra \equiv |u_1u_2\cdots u_n\ra$ term \cite{excludestate}:
\begin{equation}
	|\Psi_n\rangle = N \bigl[|c_1c_2\cdots c_n\ra -\mathcal{A}_\Omega|u_1u_2\cdots u_n\ra\bigr] \label{eq:psi1.1}.
\end{equation}
As shown in detail in the Appendix, it rewrites as
\begin{equation}
	|\Psi_n\ra=N\sum_{\alpha \subseteq \mathcal{P}(\Omega)\setminus \{\Omega\}} \mathcal{A}_\alpha \mathcal{B}_{\overline{\alpha}} \bigotimes_{i\in \alpha} |u_i\ra \otimes \bigotimes_{j\in \overline{\alpha}} |v_j\ra. \label{eq:psi1.3}
\end{equation}
The tensor product is ordered by Latin indices. The normalization constant is calculated as $N = \dfrac{1}{\sqrt{1-|\mathcal{A}_\Omega |^2}}$, as shown in the Appendix.

The state $|\Psi_n\ra$ does not contain the term $|u_1u_2\cdots u_n\ra$ in $\{|u_k\ra,|v_k\ra\}^{\otimes n}$ basis, hence it satisfies the first condition
\begin{equation}\label{eq:psicon1}
	P(\mathcal{U}_\Omega) \equiv P(U_1U_2\cdots U_n) = 0.
\end{equation}

For the next $n$ conditions, $|\Psi_n\rangle$ is examined in different bases for different particles. Particle $k^{\rm th}\in\Omega$ is measured in $\{|c_k\ra, |d_k\ra\}$, while others are measured in $\{|u\ra, |v\ra\}$. Specifically, in Eq.~\eqref{eq:psi1.1}, we substitute $|c_i\ra=A_i|u_i\ra+B_i|v_i\ra \ \forall i \neq k$ into its first term, and $|u_{k}\ra =A_k^*|c_k\ra-B_k|d_k\ra$ into the second term,
\begin{align}\label{eq:psi2}
	\notag |\Psi_n\ra &= N\Bigl[|c_1c_2\cdots c_n\ra -\mathcal{A}_\Omega |u_1u_2\cdots u_n\ra\Bigr]\\
	\notag &= N\Bigl[\bigl(A_1|u_1\ra + B_1|v_1\ra\bigr) \otimes \cdots \otimes |c_k\ra \otimes \cdots \otimes \bigl(A_n|u_n \ra + B_n|v_n\ra\bigr) \\
	\notag &  \hspace{20pt} -\mathcal{A}_{\Omega } |u_1\cdots u_{k-1}\ra \otimes \bigl(A^*_k|c_k\ra - B_k|d_k\ra\bigr) \otimes |u_{k+1}\cdots u_n\ra\Bigr] \\
	\notag &=N\Bigl[|B_k|^2 \mathcal{A}_{\overline{\kappa}} |u_1\cdots u_{k-1} c_k u_{k+1} \cdots u_n \ra \\
	& \notag \hspace{20pt} +\sum_{\alpha \in \mathcal{P}(\overline {\kappa}) \setminus \{\overline{\kappa}\}} \mathcal{A}_\alpha \mathcal{B}_{\overline{\kappa} \setminus \alpha} \bigotimes_{i\in \alpha} |u_i\ra \otimes \bigotimes_{j\in \overline{\kappa} \setminus \alpha} |v_j\ra \otimes |c_k\ra \\
	& \hspace{20pt} +
	B_k \mathcal{A}_\Omega |u_1 \cdots u_{k-1} d_k u_{k+1}\cdots u_n\ra\Bigr].
\end{align}
Here, $\kappa= \{k\}$, then $\overline{\kappa} = \Omega\setminus \kappa$. Evidently, the only term containing $d_k$ is $B_k \mathcal{A}_\Omega |u_1 \cdots d_k\cdots u_n\ra$. It means measuring $D_k=1$ infers a complete collapse of $|\Psi_n\ra$ into this substate, and all $U_i$ measurement at $i\neq k$ would yield $U_i = 1$. The second set of $n$ conditions writes as conditional probabilities
\begin{equation}\label{eq:psicon2}
	P({\mathcal{U}}_{\overline{\kappa}}| {D}_k) = 1,\ \forall k\in\Omega,
\end{equation}
which establishes an LHV-correlation that $D_k=1 \Rightarrow \mathcal{U}_{\overline{\kappa}} \equiv U_1U_2\cdots U_{k-1}  U_{k+1}\cdots U_n=1$. Every time we measure $|\Psi_n\ra$, if $D_k = 1$ then $\mathcal{U}_{\overline{\kappa}} = 1$, or $P(D_k) \le P(\mathcal{U}_{\overline{\kappa}})$. Hence, if two particles simultaneously yield $D_kD_l = 1$, $l\ne k$, then $\mathcal{U}_\Omega \equiv U_1U_2\cdots U_n = 1$. With $\overline{\lambda} = \Omega \setminus \{l\}$, the second set of conditions Eq.~\eqref{eq:psicon2} also writes
\begin{equation}\label{eq:psicon2b}
	P(D_k D_l) < P(\mathcal{U}_{\overline{\kappa} \cup \overline{\lambda}}) = P(\mathcal{U}_\Omega),\ \forall\ k,l \in \Omega, l \ne k.
\end{equation}
Therefore, measuring $|\Psi_n\ra$ should not yield $D_kD_l = 1$, $l \neq k$. However, the third set of conditions writes
\begin{align}\label{eq:psicon3}
	P({\mathcal D}_\alpha)>0, \ \forall \alpha \in \mathcal{P}(\Omega), |\alpha| \ge 2,
\end{align}
which we call the success probabilities. It results in $P(\mathcal{U}_\Omega) > 0$, which contradicts Eq.~\eqref{eq:psicon1}. We emphasize that there are more than one condition in Eq.~\eqref{eq:psicon3}.

To calculate the success probability $P({\mathcal D}_\alpha)$, the state $|\Psi_n\rangle$ from Eq.~\eqref{eq:psi1.1} is rewritten in $\{|c_k\ra, |d_k\ra\}^{\otimes n}$ basis as
\begin{align}\label{eq:psi3}
	|\Psi_n&\rangle = N\Bigl[|c_1 \cdots c_n\ra - \mathcal{A}_\Omega\Bigl(A^*_k|c_k\ra-B_k|d_k\ra\Bigr)^{\otimes n}\Bigr].
\end{align}
There are $\sum_{k=2}^{n} \binom{n}{k} = 2^n-n-1$ substates containing two or more qubits with $|\cdots d_k\cdots d_l\cdots\ra$ that satisfy Eq.\eqref{eq:psicon3}. The combined success probability is
\begin{align}\label{eq:app:non1}
	P_\text{success}=
	|\mathcal{A}_\Omega|^2-\dfrac{|\mathcal{A}_\Omega|^4}{1-|\mathcal{A}_\Omega|^2}\sum_{k = 1}^n\frac{1-|A_k|^2}{|A_k|^2}.
\end{align}
It reaches the maximum when all $|A_k|$ equals, $|A_k| = A, \ \forall k \in\Omega$. For this reason we only consider $A$ real in our data. Eq.~\eqref{eq:app:non1} becomes
\begin{align}\label{eq:Gammasame}
	P_\text{success}=A^{2n}-n\frac{A^{4n-2}(1-A^{2})}{1-A^{2n}}. 
\end{align}
Additional calculations to obtain these results are provided in detailed in the Appendix. 

These results are displayed in Fig. 1. $P_\text{success}$ as smooth curves for different system sizes $n$ are plotted as a function of $A$. In Hardy’s original setup, $A$ controls the transmission coefficients of the beamsplitters. The red curve with square dots projects $P_\text{success}$ maximal value for each $n$. It indicates that $P_\text{success}$ asymptotes 15.6\% as $n$ grows. Dots are results from a quantum simulation that runs on a virtual machine, as presented in section IV.

\section{Example: $n = 3$}
To further illustrate the above results, we calculate explicitly the Hardy's nonlocality conditions and probability for the case $n=3$. The first condition is $P(U_1U_2U_3) = 0$. The state $|\Psi_3\ra$ writes
\begin{align}
	|\Psi_3\ra & = N\Bigl[|c_1c_2c_3\ra -A_1A_2A_3 |u_1u_2u_3\ra\Bigr] \notag \\
	& = N\Bigl[\big(A_1|u_1\ra + B_1|v_1\ra\big) \big(A_2|u_2\ra + B_2|v_2\ra\big)\big(A_3|u_3\ra + B_3|v_3\ra\big) - A_1A_2A_3 |u_1u_2u_3\ra\Bigr] \notag \\
	& = N(A_{1} A_{2} B_{3} |u_{1}\ra |u_{2}\ra |v_{3}\ra + A_{1} B_{2} A_{3} |u_{1}\ra |v_{2}\ra |u_{3}\ra + A_{1} B_{2} B_{3} |u_{1}\ra |v_{2}\ra |v_{3}\ra + B_{1} A_{2} A_{3} |v_{1}\ra |u_{2}\ra |u_{3}\ra \notag \\
	& \hspace{1cm}+ B_{1} A_{2} B_{3} |v_{1}\ra |u_{2}\ra |v_{3}\ra + B_{1} B_{2} A_{3} |v_{1}\ra |v_{2}\ra |u_{3}\ra + B_{1} B_{2} B_{3} |v_{1}\ra |v_{2}\ra |v_{3}\ra).
\end{align}
To find the second condition, the state in Eq.~\eqref{eq:psi2} for the case $k=2$ is
\begin{align}
	|\Psi_3\ra &= N\Bigl[|c_1c_2c_3\ra -A_1A_2A_3 |u_1u_2u_3\ra\Bigr]\notag \\
	& = N\Bigl[\bigl(A_1|u_1\ra + B_1|v_1\ra\bigr) \otimes |c_2\ra \otimes \bigl(A_3|u_3 \ra + B_3|v_3\ra\bigr)\notag \\
	& \hspace{1 cm} -A_1A_2A_3 |u_1\ra  \otimes \bigl(A^*_2|c_2\ra - B_2|d_2\ra\bigr) \otimes 
	|u_3\ra\Bigr] \notag \\
	& = N\Bigl[A_1A_3|u_1c_2u_3\ra +B_1A_3|v_1c_2u_3\ra +A_1B_3|u_1c_2v_3\ra +B_1B_3|v_1c_2v_3\ra \notag \\
	& \hspace{1 cm}-A_1|A_2|^2A_3 |u_1c_2u_3\ra +A_1A_2A_3 B_2|u_1d_2u_3\ra\Bigr]\notag \\
	& = N\Bigl[|B_2|^2A_1A_3|u_1c_2u_3\ra	+B_1A_3|v_1c_2u_3\ra +A_1B_3|u_1c_2v_3\ra\notag \\
	& \hspace{1 cm}+B_1B_3|v_1c_2v_3\ra +A_1A_2A_3 B_2|u_1d_2u_3\ra\Bigr].
\end{align}
Only the last term $|u_1d_2u_3\ra$ contains $|d_2\ra$, thus $D_2 =1$ and $U_1U_3=1$. In another word, $P(U_1U_3|D_2) = 1$. Similarly, it is straightforward to write down the second conditions for the case  $k=1$, $P(U_2U_3|D_1) = 1$, and the case $k=3$, $P(U_1U_2|D_3) = 1$.

Finally, we verify $P(D_1D_2C_3)+P(D_1C_2D_3)+P(C_1D_2D_3)+P(D_1D_2D_3) > 0$ in
\begin{align}
	|\Psi_3\ra &= N(A_{1} A_{2} A_{3} B_{1} B_{2} B_{3} |d_{1}\ra |d_{2}\ra |d_{3}\ra - A_{1} A_{2} B_{1} B_{2} |A_{3}|^2 |d_{1}\ra |d_{2}\ra |c_{3}\ra \notag \\
	&\hspace{1cm} - A_{1} A_{3} B_{1} B_{3} |A_{2}|^2 |d_{1}\ra |c_{2}\ra |d_{3}\ra	+ A_{1} B_{1} |A_{2}|^2 |A_{3}|^2 |d_{1}\ra |c_{2}\ra |c_{3}\ra \notag \\
	&\hspace{1cm} - A_{2} A_{3} B_{2} B_{3} |A_{1}|^2 |c_{1}\ra |d_{2}\ra |d_{3}\ra	+ A_{2} B_{2} |A_{1}|^2 |A_{3}|^2 |c_{1}\ra |d_{2}\ra |c_{3}\ra \notag \\
	&\hspace{1cm} + A_{3} B_{3} |A_{1}|^2 |A_{2}|^2 |c_{1}\ra |c_{2}\ra |d_{3}\ra + (1 - |A_{1}|^2 |A_{2}|^2 |A_{3}|^2) |c_{1}\ra |c_{2}\ra |c_{3}\ra.\label{eq:n3k2}
\end{align}
Assume all $A_i$ equals and take a value 0.9: $A_{1} = A_{2} = A_{3} = A = 0.9$, we have $B_{1} = B_{2} = B_{3} = B = \sqrt{1-A^2} \approx 0.436$. The success probability for $n=3$ and $A=0.9$ is

\begin{align}
	P_\text{success} & = P(D_1D_2C_3)+P(D_1C_2D_3)+P(C_1D_2D_3)+P(D_1D_2D_3) \notag\\
	& = (|A_{1} A_{2} A_{3} B_{1} B_{2} B_{3}|^2 + | A_{1} A_{2} B_{1} B_{2} |A_{3}|^2|^2 + |A_{1} A_{3} B_{1} B_{3} |A_{2}|^2|^2 \notag\\
	&\hspace{2cm}+ |A_{2} A_{3} B_{2} B_{3} |A_{1}|^2|^2) \times \frac{1}{1-|A_{1}|^2 |A_{2}|^2 |A_{3}|^2} \notag\\
	& \approx \frac{0.9^6 \times 0.436^6+3 \times 0.9^8 \times 0.436^4}{1-0.9^6} = 0.107.
\end{align}
This result is highlighted as cyan dashed curve in Fig. 1 and also demonstrated specifically in the quantum simulation.

\section{Quantum simulation}

\begin{figure}[th]
	\begin{center}
		\includegraphics[width=0.6\textwidth]{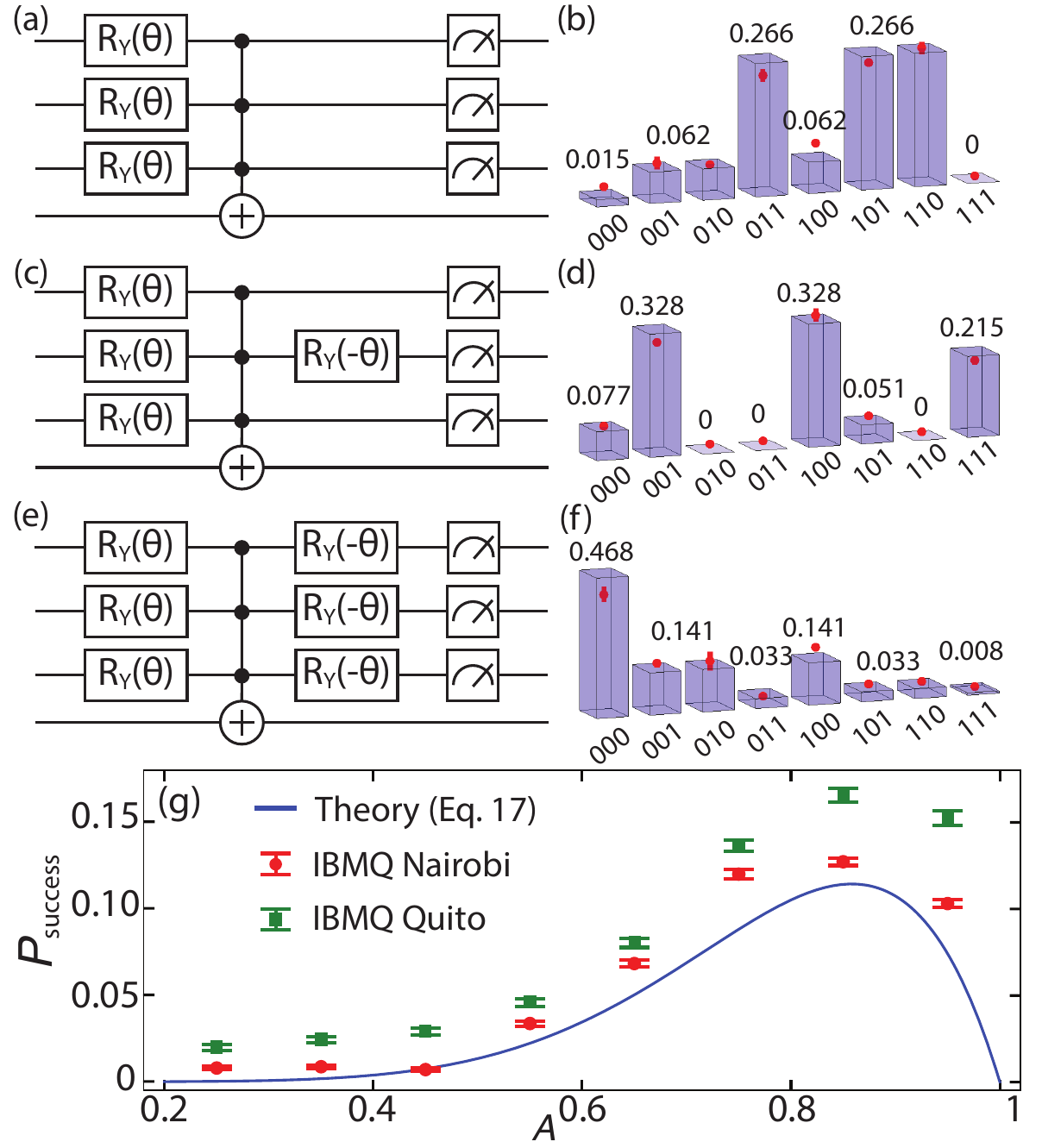}
		\caption{\textbf{Quantum simulation:} Quantum circuits for $n$ = 3 and their results executed at $A$ = 0.9, corresponding to a rotation of $\theta = 0.713\pi$ radians. Numbers and bars on the histogram are theoretical probabilities obtained from a quantum simulator, and red dots are result from IBM's Nairobi. (a, b) The first set of condition. A quantum state $|\Psi_3\rangle$ is prepared with zero probability of detecting $|u_1u_2u_3\ra$, evidently from $|111\ra$'s probability in (b). (c, d) The second set of conditions. Rotating the second qubit back to $\{|c_2\ra,|d_2\ra\}$ while leaving other qubits in $\{|u\ra,|v\ra\}$. $D_2 = 1 \Rightarrow U_1U_3=1$. (e, f) The last conditions:  applying $R_Y(-\theta)$ on all qubits. The probability for $D_kD_l=1$, $l \neq k$ is nonzero, thus establishing a Hardy-type paradox. (g) Results obtained from IBM's transmon processor Quito in green squares and Nairobi in red circles, averaging from 100 executions each with 20000 shots.}
		\label{fig:circ}
	\end{center}
\end{figure}

The above analysis can be realized on quantum hardware. Here, we implement these results to measure the nonlocality of the quantum state $|\Psi_n\rangle$ on a quantum circuit. Figure 2 illustrates our approach for $|\Psi_3\rangle$, i.e. $n$ = 3. The left column contains three quantum circuits that are consistent with the sets of Hardy's nonlocality conditions Eq.~\eqref{eq:psicon1}, \eqref{eq:psicon2}, and \eqref{eq:psicon3}. The right column shows their result for specific value $A$ = 0.9 with red dots are real data from IBM's Nairobi. Rotation gates $R_Y(\theta)$ transform each of them from XY plane to XZ plane, corresponding to a basis transformation from $\{|c\ra, |d\ra\}$ to $\{|u\ra, |v\ra\}$ basis. The transformation coefficient $A$, therefore, relates to $\theta$ as $A_k=\sin(\theta_k/2)$. State $|\Psi_n\rangle$ is constructed using post-selection scheme \cite{Aharonov2002, Lundeen2009, Yokota2009} with the help of a Toffoli gate targeting an auxiliary qubit. It is straightforward to generalize this quantum circuit to the case of an arbitrary $n$. In this case, the Toffoli gate is instead controlled by all $n$ qubits, and target on an auxiliary qubit for the post-selection scheme.
	
The first condition Eq.~\eqref{eq:psicon1} is satisfied by a state preparation with ``zero-probability substates" in it that excludes $|u_k\ra ^{\otimes n}$. This choice can manifest as a direct unitary transformation from some $|\Psi_\text{start}\ra$ into the desired state such as \cite{wu1996, Hou2021}, or a post-selection to exclude substates through joint-weak measurement \cite{Aharonov2002, Lundeen2009, Yokota2009}. Here, we employ the latter method by preparing $|\Psi_\text{start} \ra = |c_k\ra ^{\otimes n}$, then a set of rotation gates $R_Y(\theta_k)$ transforms $|\Psi_\text{start}\ra$ into $\{|u_k\ra,|v_k\ra\}^{\otimes n}$ basis. Finally, a Toffoli gate marks $|u_k\ra ^{\otimes n}$ to help post-selecting $|\Psi_n \ra$, following Eq.~\eqref{eq:psicon1}. For $n=3$, the circuit in Fig. 2a satisfies the first set of conditions by construction and yields $\la u_1 u_2 u_3|\Psi_3\ra = 0$, which is confirmed in Fig. 2b with $P(111)=0$.

The second set of conditions Eq.~\eqref{eq:psicon2} is tested by measuring the $k^\text{th}$ qubit in $\{|c_k\ra,|d_k\ra\}$ basis, while leaving others in their $\{|u_i\ra,|v_i\ra \}$ bases. This is done on the circuit by applying a single $R_Y(-\theta_k)$ on the $k^\text{th}$ qubit. In Fig. \ref{fig:circ}c, if the measurement yields $D_2$ = 1, it ensures $U_1 = U_3 = 1$. This means $P(U_1 D_2 U_3) \equiv P(111) = 0.215 > 0$, while $P(V_1 D_2 U_3) \equiv P(011) = 0$, $P(U_1 D_2 V_3) \equiv P(110) = 0$, and $P(V_1 D_2 V_3) \equiv P(010) = 0$, as shown in Fig. \ref{fig:circ}d. It is straightforward to verify that $R_Y(-\theta_k)$ has a similar effect when applied on $|q_1\ra$ or $|q_3\ra$, or the case of generalized $n$. In other words, the same correlation is found when measuring $D_1U_2U_3$ or $U_1U_2D_3$.
	
The third set of conditions Eq.~\eqref{eq:psicon3} is a nonzero total probability of measurements that contradicts previous conditions, generally tested in $\{|c_k\ra,|d_k\ra\}^{\otimes n}$. To check, we apply $R_Y(-\theta_k)$ on all qubits and measure them in $\{|c_k\ra,|d_k\ra\}^{\otimes n}$ basis. For $n$=3, these states are $|d_1 d_2 c_3\ra$, $|d_1 c_2 d_3\ra$, $|c_1 d_2 d_3\ra$, and $|d_1 d_2 d_3\ra$. In our quantum circuit, the corresponding states are $|110\ra$, $|101\ra$, $|011\ra$, and $|111\ra$, which yield a combined probability of 0.107 as shown in Fig. \ref{fig:circ}f, in agreement with theoretical result in Eq. \eqref{eq:Gammasame} for $A$ = 0.9 and $n$ = 3. Henceforth, all conditions are satisfied, and a Hardy-type paradox is established.

\begin{table*}
	\begin{center}		
	\begin{tabular}{|c|c|c|c|c|c|c|c|}\hline 
		Device & Qubit & $T_1$ & $T_2$ & CNOT & Readout & Single gate & Connectivity \\
		(mm/yyyy) & & ($\mu$s) & ($\mu$s) & error (E-2) & error (E-2) & error (E-4) & \\
		\hline\hline
		Quito & Q0 & 72.54 & 157.41 & 0-1:1.31 & 3.51 & 3.92 & 0-1 \\
		04/2022 & Q1 & 81.5 & 121.88 & 1-3:1.14 & 2.10 & 3.30 & 1-0, 1-2, 1-3 \\
		& Q2 & 108.39 & 114.18 & 2-1:0.77 & 8.70 & 5.19 & 2-1   \\
		& Q3 & 109.36 & 20.86 & 3-4:2.13 & 2.81 & 9.28 & 3-1, 3-4   \\
		& Q4 & 69.45 & 62.78 &  & 2.91 & 4.26 & 4-3   \\
        \hline
		Nairobi & Q0 & 91.56 & 37.19 & 0-1:0.98 & 4.53 & 2.96 & 0-1   \\
		07/2022& Q1 & 100.08 & 69.6 & 1-3:0.75 & 2.70 & 3.49 & 1-0, 1-2, 1-3   \\
		& Q2 & 95.6 & 134.33 & 2-1:0.93 & 3.58 & 3.64 & 2-1   \\
		& Q3 & 152.05 & 37.16 & 3-5:1.46 & 2.65 & 3.54 & 3-1, 3-5   \\
		& Q4 & 118.72 & 90.71 & 4-5:0.50 & 2.69 & 2.93 & 4-5   \\
		& Q5 & 162.52 & 22.81 & 5-6:0.87 & 2.96 & 2.82 & 5-3, 5-4, 5-6   \\
		& Q6 & 140.26 & 90.9 &  & 3.36 & 2.23 & 6-5   \\
		\hline
	\end{tabular}
	\end{center}
	\caption{Calibration of IBM devices, as recorded at the time of the runs. The basis single gates are $ID, RZ, SX, X$. Columns showing lifetime $T_1$, decoherence time $T_2$, gate errors, and readout errors. Connectivity denotes the layout of the chip. The calibration files can be found on Github \cite{github}.}
	\label{tab:calibration}
\end{table*}

To demonstrate the practical aspect of this simulation, we execute the quantum circuit in Fig. 2e on IBM's quantum computers Nairobi and Quito \cite{github} and obtain results shown on Fig. 2g, averaging from 2 million shots each, equally divided into 100 different runs. The detail parameters of these machines are given in table 1. Apparently, the systematic errors on Quito are slightly larger than those on Nairobi. These are the best results we obtain from IBM's transmon processors. Any demonstration for $n > 3$ yields result with big errors, as compared to IBM's own QASM simulator. 

Obtaining identical result for $P_\text{success}$, but our analytical calculation and the quantum simulation are different approaches. In the first method, we start with a quantum state and analytically derive its conditions and a formula for $P_\text{success}$ as in Eq.~\eqref{eq:Gammasame}. It is a theoretical model of nonlocality. In the second part, a quantum state is evolved in a specific quantum circuit, and the success probability is obtained from a measurement. It is an experimental protocol to probe nonlocality that can be implemented on real quantum hardware. For example, our circuit on an optical quantum computer would produce $n$ intertwined Mach–Zehnder interferometers. As argued above and in Fig. 2, the quantum circuit is equivalent to Hardy's nonlocality conditions Eq. \eqref{eq:psicon1}, \eqref{eq:psicon2}, \eqref{eq:psicon3}. As shown in Fig. 1, the overlapping between dots and solid lines confirms the indistinguishable results from two distinct approaches. The simulation, however, limits to small $n$ due to exponential growth in computing memory. 

\section{Nonlocality and entanglement}

\begin{figure}[htb]
	\centering
	\includegraphics[width=0.8\textwidth]{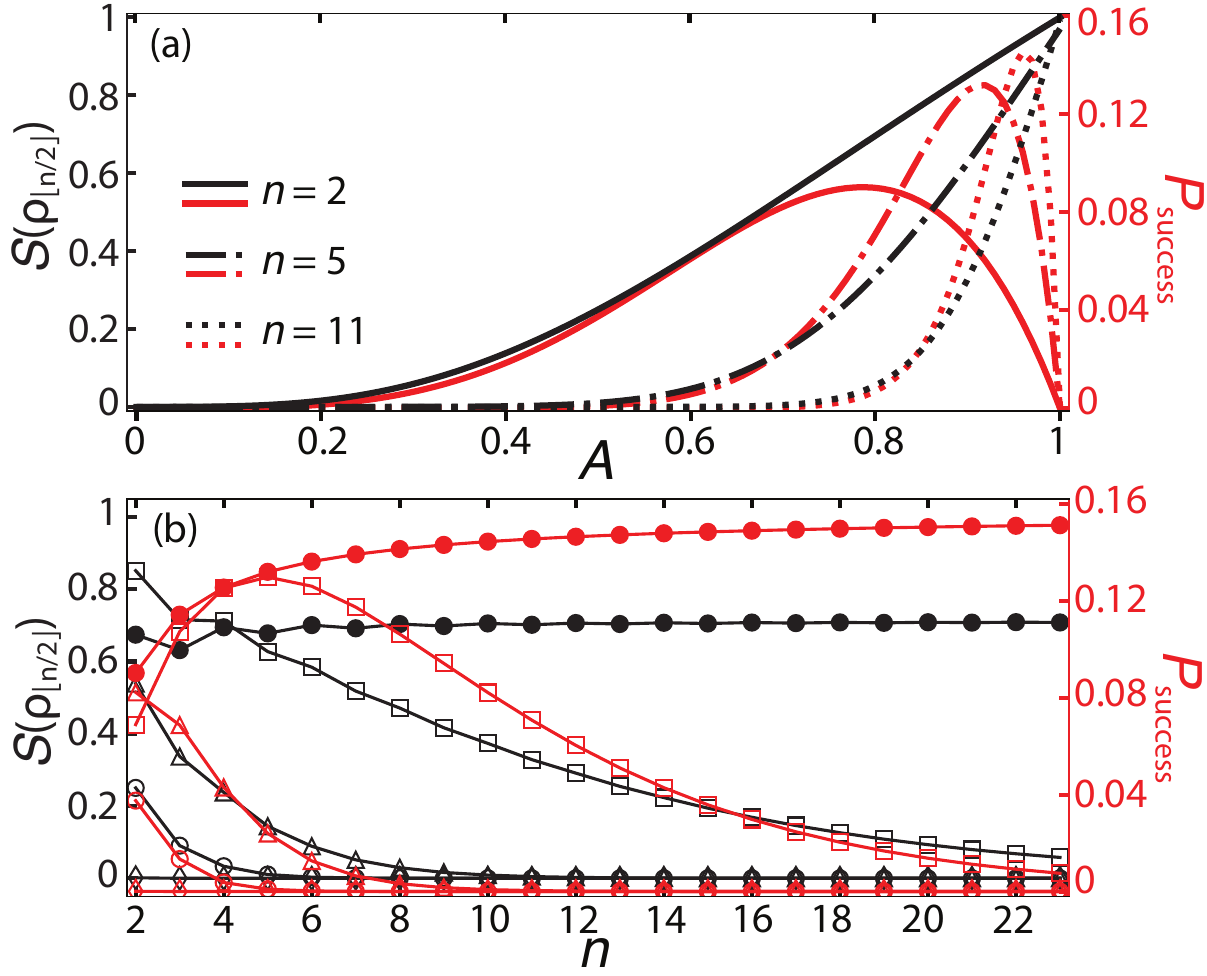}
	\caption{\textbf{Entanglement entropy:} von Neumann entropy $S(\rho_{\lfloor n/2 \rfloor})$ calculated at half-chain bipartition is shown on the left axis in black. In direct comparison, $P_\text{success}$ per Eq.~\eqref{eq:Gammasame} is shown on the right axis in red (gray). (a) $S(\rho_{\lfloor n/2 \rfloor})$ and $P_\text{success}$ as a function of the transformation coefficient $A$ at $n$ = 2, 5, and 11. (b) $S(\rho_{\lfloor n/2 \rfloor})$ and $P_\text{success}$ as a function of $n$ at $A$ = 0.1 (open diamonds), 0.5 (open circles), 0.7 (open triangles), and 0.9 (open squares). The solid circles are data obtained at the optimum $A$, i.e. at the maximum value of $P_\text{success}$.}
	\label{fig:entanglement}
\end{figure}

The connection between entanglement and quantum nonlocality is one of contentious nature \cite{Brunner2005, Junge2011, Liang2011, Vidick2011}. To further study $|\Psi_n\rangle$, its entanglement is calculated under half-chain bipartition using von Neumann entropy. Let $\rho = |\Psi_n\ra\la \Psi_n|$ be the density matrix of $|\Psi_n\ra$ on the Hilbert space $\mathcal{H} = \mathcal{H}_{\mathcal{A}} \otimes \mathcal{H}_{\mathcal{B}}$. Then, $\mathcal{A}$'s reduced density matrix is $\rho_{\mathcal{A}} = \operatorname{tr}_{\mathcal{B}}(\rho)$. The von Neumann entropy at half-chain bipartition writes
\begin{equation}
	S(\rho_{\lfloor n/2 \rfloor})=-\operatorname{tr}(\rho_{\lfloor n/2 \rfloor} \log_2 \rho_{\lfloor n/2 \rfloor}),
\end{equation}
with $\lfloor n/2 \rfloor$ denotes the floor of $n/2$. These two quantities $P_\text{success}$ and $S(\rho_{\lfloor n/2 \rfloor})$ are compared in Fig.~\ref{fig:entanglement} as a function of the transformation coefficient $A$ and the size $n$. Apparently, their behavior are completely different. While $P_\text{success}$ increases to a maximum and then decreases sharply, $S(\rho_{\lfloor n/2 \rfloor})$ grows monotonically. It implies that high entanglement does not always correlate to high nonlocality \cite{Junge2011, Brunner2005, Vidick2011}. Another trend is observed when $S(\rho_{\lfloor n/2 \rfloor})$ and $P_\text{success}$ are plotted as a function of $n$ for different $A$, shown in Fig.~\ref{fig:entanglement}b. Only for optimal $A$, $P_\text{success}$ and $S(\rho_{\lfloor n/2 \rfloor})$ approach a constant as $n$ grows, as shown by solid circles in Fig.~\ref{fig:entanglement}b. Parallel analyses under a different bipartition for one-versus-the-rest of the system, and using the positive partial transpose (PPT), also show monotonic increases, as shown in detail in Appendix C.

\begin{figure}[t]
	\centering
	\includegraphics[width=0.8\textwidth]{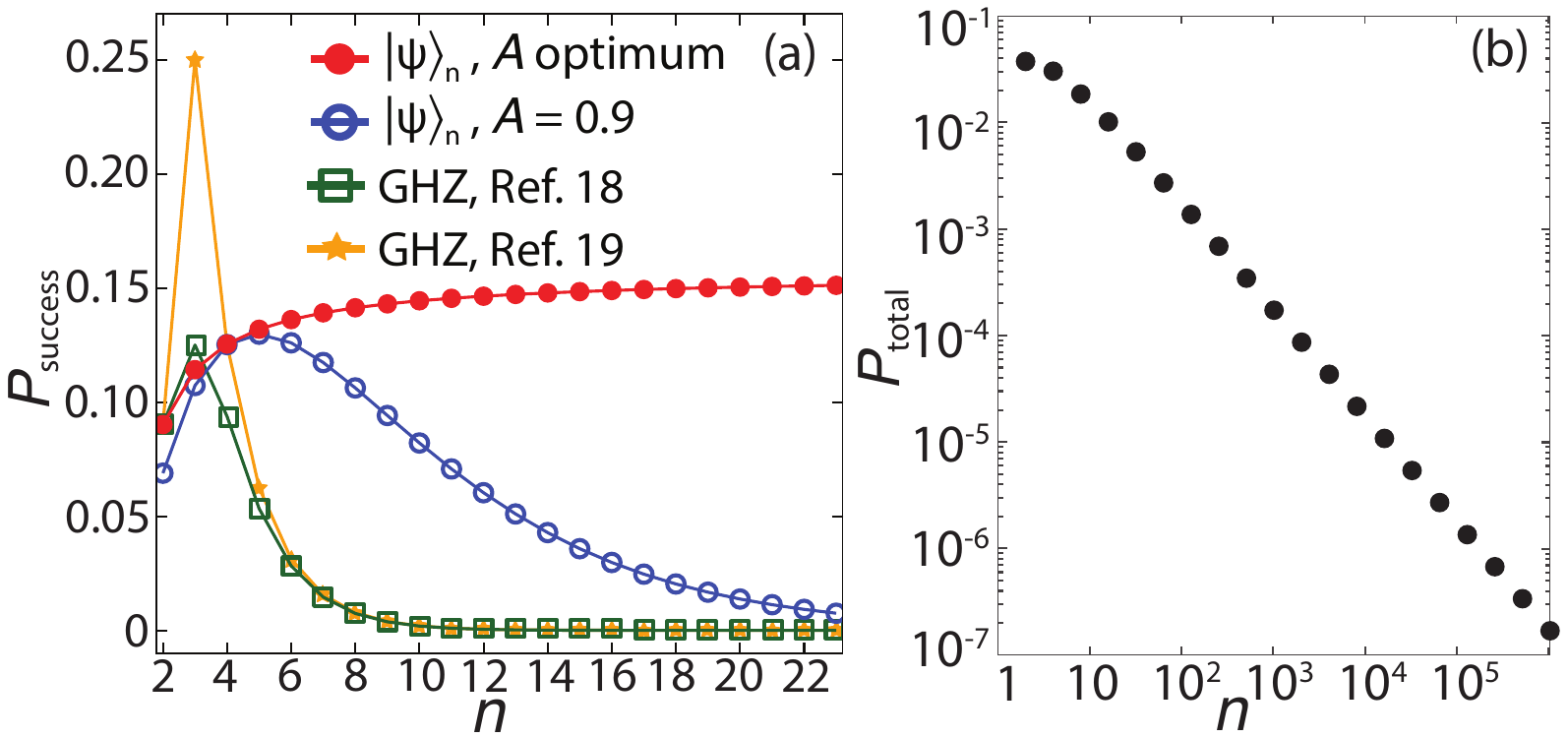}
	\caption{\textbf{Nonlocality at the macroscopic scale:} (a) $P_\text{success}$ for $|\Psi_n\ra$ per Eq.~\eqref{eq:Gammasame} at two values: $A$ = 0.9 in blue open circles and $A$ optimum in red solid circles are directly compared to results for generalized GHZ state in Eq. (28),  Ref.~\cite{cereceda2004} in green open squares and Eq. (2), Ref.~\cite{jiang2018} in yellow solid stars. (b) The total success probability $P_\text{total}$ quickly reduces to zero as $n$ grows in a log-log plot.}
\label{fig:summary}
\end{figure}

\section{Discussion and Conclusion.}
Apparently, $|\Psi_n\ra$ defined in Eq.~\eqref{eq:psi1.1} is not a common state such as GHZ, W, or Dicke. Indeed, our choice was inspired by the symmetry in the quantum circuit that simulates original Hardy's paradox in our previous work \cite{Tran2022}. Here, the generalized quantum circuit is a natural extension from Fig. 4b in Ref.~\cite{Tran2022} where the Toffoli gate is controlled by all $n$ qubits. 

As $n$ grows, $P_\text{success}$ quickly reduces to zero at fixed $A$, as shown in Fig.~\ref{fig:summary}a for a representative value $A$ = 0.9. This observation matches with previous work on generalized GHZ states. At $n$ = 2, $P_\text{success}$ =9\% because they are all equal to the original Hardy's result. The fast reduction of $P_\text{success}$ at large $n$ in Ref. \cite{cereceda2004,jiang2018} might relates to the nature of GHZ states. Although maximally entangled, GHZ states do not necessarily reach maximum in nonlocality \cite{Brunner2005, Junge2011, Liang2011, Vidick2011}. There is an optimum value of $A = |A_k| \ \forall k \in\Omega$ that $P_\text{success}$ is maximized whereas the entanglement entropy $S(\rho_{\lfloor n/2 \rfloor})$ behaves monotonically.

However, it is difficult to detect this success probability of 15.6\% at the macrosopic scale. The total success probability of an $n$-particle system can be calculated from the area of the $P_\text{success}(A)$ curve $P_{\rm{total}}=\int_0^1 P_\text{success}(A) dA.$ As seen in Fig.~\ref{fig:summary}b, $P_{\rm{total}}$ quickly reduces to zero as $n$ grows. This is also evident from Fig. 1, as the probability curves become narrower with larger $n$. A small change in the transformation coefficient $A$ leads to a sudden drop in $P_\text{success}$. The success probability $P_\text{success}$ is a measure of how likely one can see Hardy's nonlocality. In our scheme, instead of the $P_\text{success}$ dropping as $n$ grows, it persists, while becoming increasingly sensitive to fluctuation in transformation coefficient parameters $A$. In any case, difficulty in demonstration does not mean the absence of nonlocal features of quantum systems.

In summary, we have extended the original Hardy's paradox to a general case and obtained a nonlocality with probability approaching 15.6\% as the  system size grows. The Hardy's nonlocality conditions and success probability are derived analytically. A quantum simulation is proposed that matches well to the theory, especially when tested on real quantum computers.

{\it Acknowledgement}--- HQN would like to thank Dr. Nguyen Hoang Hai and Dr. Nguyen The Toan for helpful discussions.

\bibliographystyle{apsrev4-2}
\bibliography{nonlocbib}

\begin{thebibliography}{56}%
\makeatletter
\providecommand \@ifxundefined [1]{%
 \@ifx{#1\undefined}
}%
\providecommand \@ifnum [1]{%
 \ifnum #1\expandafter \@firstoftwo
 \else \expandafter \@secondoftwo
 \fi
}%
\providecommand \@ifx [1]{%
 \ifx #1\expandafter \@firstoftwo
 \else \expandafter \@secondoftwo
 \fi
}%
\providecommand \natexlab [1]{#1}%
\providecommand \enquote  [1]{``#1''}%
\providecommand \bibnamefont  [1]{#1}%
\providecommand \bibfnamefont [1]{#1}%
\providecommand \citenamefont [1]{#1}%
\providecommand \href@noop [0]{\@secondoftwo}%
\providecommand \href [0]{\begingroup \@sanitize@url \@href}%
\providecommand \@href[1]{\@@startlink{#1}\@@href}%
\providecommand \@@href[1]{\endgroup#1\@@endlink}%
\providecommand \@sanitize@url [0]{\catcode `\\12\catcode `\$12\catcode
  `\&12\catcode `\#12\catcode `\^12\catcode `\_12\catcode `\%12\relax}%
\providecommand \@@startlink[1]{}%
\providecommand \@@endlink[0]{}%
\providecommand \url  [0]{\begingroup\@sanitize@url \@url }%
\providecommand \@url [1]{\endgroup\@href {#1}{\urlprefix }}%
\providecommand \urlprefix  [0]{URL }%
\providecommand \Eprint [0]{\href }%
\providecommand \doibase [0]{https://doi.org/}%
\providecommand \selectlanguage [0]{\@gobble}%
\providecommand \bibinfo  [0]{\@secondoftwo}%
\providecommand \bibfield  [0]{\@secondoftwo}%
\providecommand \translation [1]{[#1]}%
\providecommand \BibitemOpen [0]{}%
\providecommand \bibitemStop [0]{}%
\providecommand \bibitemNoStop [0]{.\EOS\space}%
\providecommand \EOS [0]{\spacefactor3000\relax}%
\providecommand \BibitemShut  [1]{\csname bibitem#1\endcsname}%
\let\auto@bib@innerbib\@empty
\bibitem [{\citenamefont {Hardy}(1992)}]{hardy1992}%
  \BibitemOpen
  \bibfield  {author} {\bibinfo {author} {\bibfnamefont {L.}~\bibnamefont
  {Hardy}},\ }\href@noop {} {\bibfield  {journal} {\bibinfo  {journal}
  {Physical Review Letters}\ }\textbf {\bibinfo {volume} {68}},\ \bibinfo
  {pages} {2981} (\bibinfo {year} {1992})}\BibitemShut {NoStop}%
\bibitem [{\citenamefont {Hardy}(1993)}]{hardy1993}%
  \BibitemOpen
  \bibfield  {author} {\bibinfo {author} {\bibfnamefont {L.}~\bibnamefont
  {Hardy}},\ }\href@noop {} {\bibfield  {journal} {\bibinfo  {journal}
  {Physical Review Letters}\ }\textbf {\bibinfo {volume} {71}},\ \bibinfo
  {pages} {1665} (\bibinfo {year} {1993})}\BibitemShut {NoStop}%
\bibitem [{\citenamefont {Kar}(1997{\natexlab{a}})}]{kar1997_nonlocality_term}%
  \BibitemOpen
  \bibfield  {author} {\bibinfo {author} {\bibfnamefont {G.}~\bibnamefont
  {Kar}},\ }\href@noop {} {\bibfield  {journal} {\bibinfo  {journal} {Journal
  of Physics A: Mathematical and General}\ }\textbf {\bibinfo {volume} {30}},\
  \bibinfo {pages} {L217} (\bibinfo {year} {1997}{\natexlab{a}})}\BibitemShut
  {NoStop}%
\bibitem [{\citenamefont {Ghosh}\ \emph {et~al.}(1998)\citenamefont {Ghosh},
  \citenamefont {Kar},\ and\ \citenamefont {Sarkar}}]{ghosh1998}%
  \BibitemOpen
  \bibfield  {author} {\bibinfo {author} {\bibfnamefont {S.}~\bibnamefont
  {Ghosh}}, \bibinfo {author} {\bibfnamefont {G.}~\bibnamefont {Kar}},\ and\
  \bibinfo {author} {\bibfnamefont {D.}~\bibnamefont {Sarkar}},\ }\href@noop {}
  {\bibfield  {journal} {\bibinfo  {journal} {Physics Letters A}\ }\textbf
  {\bibinfo {volume} {243}},\ \bibinfo {pages} {249} (\bibinfo {year}
  {1998})}\BibitemShut {NoStop}%
\bibitem [{\citenamefont {Cabello}\ \emph {et~al.}(2013)\citenamefont
  {Cabello}, \citenamefont {Badziag}, \citenamefont {Cunha},\ and\
  \citenamefont {Bourennane}}]{Cabello2013}%
  \BibitemOpen
  \bibfield  {author} {\bibinfo {author} {\bibfnamefont {A.}~\bibnamefont
  {Cabello}}, \bibinfo {author} {\bibfnamefont {P.}~\bibnamefont {Badziag}},
  \bibinfo {author} {\bibfnamefont {M.~T.}\ \bibnamefont {Cunha}},\ and\
  \bibinfo {author} {\bibfnamefont {M.}~\bibnamefont {Bourennane}},\
  }\href@noop {} {\bibfield  {journal} {\bibinfo  {journal} {Physical review
  letters}\ }\textbf {\bibinfo {volume} {111}},\ \bibinfo {pages} {180404}
  (\bibinfo {year} {2013})}\BibitemShut {NoStop}%
\bibitem [{\citenamefont {Popescu}\ and\ \citenamefont
  {Rohrlich}(1992)}]{Popescu1992}%
  \BibitemOpen
  \bibfield  {author} {\bibinfo {author} {\bibfnamefont {S.}~\bibnamefont
  {Popescu}}\ and\ \bibinfo {author} {\bibfnamefont {D.}~\bibnamefont
  {Rohrlich}},\ }\href@noop {} {\bibfield  {journal} {\bibinfo  {journal}
  {Physics Letters A}\ }\textbf {\bibinfo {volume} {166}},\ \bibinfo {pages}
  {293} (\bibinfo {year} {1992})}\BibitemShut {NoStop}%
\bibitem [{\citenamefont {Cabello}(2000)}]{cabello1999}%
  \BibitemOpen
  \bibfield  {author} {\bibinfo {author} {\bibfnamefont {A.}~\bibnamefont
  {Cabello}},\ }\href@noop {} {\bibfield  {journal} {\bibinfo  {journal}
  {Physical Review A}\ }\textbf {\bibinfo {volume} {61}},\ \bibinfo {pages}
  {022119} (\bibinfo {year} {2000})}\BibitemShut {NoStop}%
\bibitem [{\citenamefont {{\.Z}ukowski}\ \emph {et~al.}(2002)\citenamefont
  {{\.Z}ukowski}, \citenamefont {Brukner}, \citenamefont {Laskowski},\ and\
  \citenamefont {Wie{\'s}niak}}]{Zukowski2002}%
  \BibitemOpen
  \bibfield  {author} {\bibinfo {author} {\bibfnamefont {M.}~\bibnamefont
  {{\.Z}ukowski}}, \bibinfo {author} {\bibfnamefont {{\v{C}}.}~\bibnamefont
  {Brukner}}, \bibinfo {author} {\bibfnamefont {W.}~\bibnamefont {Laskowski}},\
  and\ \bibinfo {author} {\bibfnamefont {M.}~\bibnamefont {Wie{\'s}niak}},\
  }\href@noop {} {\bibfield  {journal} {\bibinfo  {journal} {Physical review
  letters}\ }\textbf {\bibinfo {volume} {88}},\ \bibinfo {pages} {210402}
  (\bibinfo {year} {2002})}\BibitemShut {NoStop}%
\bibitem [{\citenamefont {Yang}\ \emph {et~al.}(2019)\citenamefont {Yang},
  \citenamefont {Meng}, \citenamefont {Zhou}, \citenamefont {Xu}, \citenamefont
  {Xiao}, \citenamefont {Sun}, \citenamefont {Chen}, \citenamefont {Xu},
  \citenamefont {Li},\ and\ \citenamefont {Guo}}]{YangPRA19}%
  \BibitemOpen
  \bibfield  {author} {\bibinfo {author} {\bibfnamefont {M.}~\bibnamefont
  {Yang}}, \bibinfo {author} {\bibfnamefont {H.-X.}\ \bibnamefont {Meng}},
  \bibinfo {author} {\bibfnamefont {J.}~\bibnamefont {Zhou}}, \bibinfo {author}
  {\bibfnamefont {Z.-P.}\ \bibnamefont {Xu}}, \bibinfo {author} {\bibfnamefont
  {Y.}~\bibnamefont {Xiao}}, \bibinfo {author} {\bibfnamefont {K.}~\bibnamefont
  {Sun}}, \bibinfo {author} {\bibfnamefont {J.-L.}\ \bibnamefont {Chen}},
  \bibinfo {author} {\bibfnamefont {J.-S.}\ \bibnamefont {Xu}}, \bibinfo
  {author} {\bibfnamefont {C.-F.}\ \bibnamefont {Li}},\ and\ \bibinfo {author}
  {\bibfnamefont {G.-C.}\ \bibnamefont {Guo}},\ }\href@noop {} {\bibfield
  {journal} {\bibinfo  {journal} {Physical Review A}\ }\textbf {\bibinfo
  {volume} {99}},\ \bibinfo {pages} {032103} (\bibinfo {year}
  {2019})}\BibitemShut {NoStop}%
\bibitem [{\citenamefont {Vallone}\ \emph {et~al.}(2011)\citenamefont
  {Vallone}, \citenamefont {Gianani}, \citenamefont {Inostroza}, \citenamefont
  {Saavedra}, \citenamefont {Lima}, \citenamefont {Cabello},\ and\
  \citenamefont {Mataloni}}]{VallonePRA11}%
  \BibitemOpen
  \bibfield  {author} {\bibinfo {author} {\bibfnamefont {G.}~\bibnamefont
  {Vallone}}, \bibinfo {author} {\bibfnamefont {I.}~\bibnamefont {Gianani}},
  \bibinfo {author} {\bibfnamefont {E.~B.}\ \bibnamefont {Inostroza}}, \bibinfo
  {author} {\bibfnamefont {C.}~\bibnamefont {Saavedra}}, \bibinfo {author}
  {\bibfnamefont {G.}~\bibnamefont {Lima}}, \bibinfo {author} {\bibfnamefont
  {A.}~\bibnamefont {Cabello}},\ and\ \bibinfo {author} {\bibfnamefont
  {P.}~\bibnamefont {Mataloni}},\ }\href@noop {} {\bibfield  {journal}
  {\bibinfo  {journal} {Physical Review A}\ }\textbf {\bibinfo {volume} {83}},\
  \bibinfo {pages} {042105} (\bibinfo {year} {2011})}\BibitemShut {NoStop}%
\bibitem [{\citenamefont {Jordan}(1994)}]{JordanPRA94}%
  \BibitemOpen
  \bibfield  {author} {\bibinfo {author} {\bibfnamefont {T.~F.}\ \bibnamefont
  {Jordan}},\ }\href@noop {} {\bibfield  {journal} {\bibinfo  {journal}
  {Physical Review A}\ }\textbf {\bibinfo {volume} {50}},\ \bibinfo {pages}
  {62} (\bibinfo {year} {1994})}\BibitemShut {NoStop}%
\bibitem [{\citenamefont {Goldstein}(1994)}]{goldstein1994}%
  \BibitemOpen
  \bibfield  {author} {\bibinfo {author} {\bibfnamefont {S.}~\bibnamefont
  {Goldstein}},\ }\href@noop {} {\bibfield  {journal} {\bibinfo  {journal}
  {Physical review letters}\ }\textbf {\bibinfo {volume} {72}},\ \bibinfo
  {pages} {1951} (\bibinfo {year} {1994})}\BibitemShut {NoStop}%
\bibitem [{\citenamefont {Wu}\ and\ \citenamefont {Xie}(1996)}]{wu1996}%
  \BibitemOpen
  \bibfield  {author} {\bibinfo {author} {\bibfnamefont {X.-h.}\ \bibnamefont
  {Wu}}\ and\ \bibinfo {author} {\bibfnamefont {R.-h.}\ \bibnamefont {Xie}},\
  }\href@noop {} {\bibfield  {journal} {\bibinfo  {journal} {Physics Letters
  A}\ }\textbf {\bibinfo {volume} {211}},\ \bibinfo {pages} {129} (\bibinfo
  {year} {1996})}\BibitemShut {NoStop}%
\bibitem [{\citenamefont {Wu}\ \emph {et~al.}(2000)\citenamefont {Wu},
  \citenamefont {Zong},\ and\ \citenamefont {Pang}}]{wu2000}%
  \BibitemOpen
  \bibfield  {author} {\bibinfo {author} {\bibfnamefont {X.-H.}\ \bibnamefont
  {Wu}}, \bibinfo {author} {\bibfnamefont {H.-S.}\ \bibnamefont {Zong}},\ and\
  \bibinfo {author} {\bibfnamefont {H.-R.}\ \bibnamefont {Pang}},\ }\href@noop
  {} {\bibfield  {journal} {\bibinfo  {journal} {Physics Letters A}\ }\textbf
  {\bibinfo {volume} {276}},\ \bibinfo {pages} {221} (\bibinfo {year}
  {2000})}\BibitemShut {NoStop}%
\bibitem [{\citenamefont {Kar}(1997{\natexlab{b}})}]{kar1997}%
  \BibitemOpen
  \bibfield  {author} {\bibinfo {author} {\bibfnamefont {G.}~\bibnamefont
  {Kar}},\ }\href@noop {} {\bibfield  {journal} {\bibinfo  {journal} {Physical
  Review A}\ }\textbf {\bibinfo {volume} {56}},\ \bibinfo {pages} {1023}
  (\bibinfo {year} {1997}{\natexlab{b}})}\BibitemShut {NoStop}%
\bibitem [{\citenamefont {Ghosh}\ and\ \citenamefont {Roy}(2010)}]{ghosh2010}%
  \BibitemOpen
  \bibfield  {author} {\bibinfo {author} {\bibfnamefont {S.}~\bibnamefont
  {Ghosh}}\ and\ \bibinfo {author} {\bibfnamefont {S.~M.}\ \bibnamefont
  {Roy}},\ }\href@noop {} {\bibfield  {journal} {\bibinfo  {journal} {Journal
  of mathematical physics}\ }\textbf {\bibinfo {volume} {51}},\ \bibinfo
  {pages} {122204} (\bibinfo {year} {2010})}\BibitemShut {NoStop}%
\bibitem [{\citenamefont {Pagonis}\ and\ \citenamefont
  {Clifton}(1992)}]{pagonis1992}%
  \BibitemOpen
  \bibfield  {author} {\bibinfo {author} {\bibfnamefont {C.}~\bibnamefont
  {Pagonis}}\ and\ \bibinfo {author} {\bibfnamefont {R.}~\bibnamefont
  {Clifton}},\ }\href@noop {} {\bibfield  {journal} {\bibinfo  {journal}
  {Physics Letters A}\ }\textbf {\bibinfo {volume} {168}},\ \bibinfo {pages}
  {100} (\bibinfo {year} {1992})}\BibitemShut {NoStop}%
\bibitem [{\citenamefont {Cereceda}(2004)}]{cereceda2004}%
  \BibitemOpen
  \bibfield  {author} {\bibinfo {author} {\bibfnamefont {J.~L.}\ \bibnamefont
  {Cereceda}},\ }\href@noop {} {\bibfield  {journal} {\bibinfo  {journal}
  {Physics Letters A}\ }\textbf {\bibinfo {volume} {327}},\ \bibinfo {pages}
  {433} (\bibinfo {year} {2004})}\BibitemShut {NoStop}%
\bibitem [{\citenamefont {Jiang}\ \emph {et~al.}(2018)\citenamefont {Jiang},
  \citenamefont {Xu}, \citenamefont {Su}, \citenamefont {Pati},\ and\
  \citenamefont {Chen}}]{jiang2018}%
  \BibitemOpen
  \bibfield  {author} {\bibinfo {author} {\bibfnamefont {S.-H.}\ \bibnamefont
  {Jiang}}, \bibinfo {author} {\bibfnamefont {Z.-P.}\ \bibnamefont {Xu}},
  \bibinfo {author} {\bibfnamefont {H.-Y.}\ \bibnamefont {Su}}, \bibinfo
  {author} {\bibfnamefont {A.~K.}\ \bibnamefont {Pati}},\ and\ \bibinfo
  {author} {\bibfnamefont {J.-L.}\ \bibnamefont {Chen}},\ }\href@noop {}
  {\bibfield  {journal} {\bibinfo  {journal} {Physical review letters}\
  }\textbf {\bibinfo {volume} {120}},\ \bibinfo {pages} {050403} (\bibinfo
  {year} {2018})}\BibitemShut {NoStop}%
\bibitem [{\citenamefont {Cabello}\ \emph {et~al.}(2008)\citenamefont
  {Cabello}, \citenamefont {G{\"u}hne}, \citenamefont {Moreno},\ and\
  \citenamefont {Rodr{\'\i}guez}}]{Cabello2008}%
  \BibitemOpen
  \bibfield  {author} {\bibinfo {author} {\bibfnamefont {A.}~\bibnamefont
  {Cabello}}, \bibinfo {author} {\bibfnamefont {O.}~\bibnamefont {G{\"u}hne}},
  \bibinfo {author} {\bibfnamefont {P.}~\bibnamefont {Moreno}},\ and\ \bibinfo
  {author} {\bibfnamefont {D.}~\bibnamefont {Rodr{\'\i}guez}},\ }\href@noop {}
  {\bibfield  {journal} {\bibinfo  {journal} {Laser physics}\ }\textbf
  {\bibinfo {volume} {18}},\ \bibinfo {pages} {335} (\bibinfo {year}
  {2008})}\BibitemShut {NoStop}%
\bibitem [{\citenamefont {Gachechiladze}\ \emph {et~al.}(2016)\citenamefont
  {Gachechiladze}, \citenamefont {Budroni},\ and\ \citenamefont
  {G{\"u}hne}}]{Gachechiladze2015}%
  \BibitemOpen
  \bibfield  {author} {\bibinfo {author} {\bibfnamefont {M.}~\bibnamefont
  {Gachechiladze}}, \bibinfo {author} {\bibfnamefont {C.}~\bibnamefont
  {Budroni}},\ and\ \bibinfo {author} {\bibfnamefont {O.}~\bibnamefont
  {G{\"u}hne}},\ }\href@noop {} {\bibfield  {journal} {\bibinfo  {journal}
  {Physical review letters}\ }\textbf {\bibinfo {volume} {116}},\ \bibinfo
  {pages} {070401} (\bibinfo {year} {2016})}\BibitemShut {NoStop}%
\bibitem [{\citenamefont {Wang}\ \emph {et~al.}(2012)\citenamefont {Wang},
  \citenamefont {Markham} \emph {et~al.}}]{Wang2012}%
  \BibitemOpen
  \bibfield  {author} {\bibinfo {author} {\bibfnamefont {Z.}~\bibnamefont
  {Wang}}, \bibinfo {author} {\bibfnamefont {D.}~\bibnamefont {Markham}}, \emph
  {et~al.},\ }\href@noop {} {\bibfield  {journal} {\bibinfo  {journal}
  {Physical review letters}\ }\textbf {\bibinfo {volume} {108}},\ \bibinfo
  {pages} {210407} (\bibinfo {year} {2012})}\BibitemShut {NoStop}%
\bibitem [{\citenamefont {Barnea}\ \emph {et~al.}(2015)\citenamefont {Barnea},
  \citenamefont {P{\"u}tz}, \citenamefont {Brask}, \citenamefont {Brunner},
  \citenamefont {Gisin},\ and\ \citenamefont {Liang}}]{Barnea2015}%
  \BibitemOpen
  \bibfield  {author} {\bibinfo {author} {\bibfnamefont {T.~J.}\ \bibnamefont
  {Barnea}}, \bibinfo {author} {\bibfnamefont {G.}~\bibnamefont {P{\"u}tz}},
  \bibinfo {author} {\bibfnamefont {J.~B.}\ \bibnamefont {Brask}}, \bibinfo
  {author} {\bibfnamefont {N.}~\bibnamefont {Brunner}}, \bibinfo {author}
  {\bibfnamefont {N.}~\bibnamefont {Gisin}},\ and\ \bibinfo {author}
  {\bibfnamefont {Y.-C.}\ \bibnamefont {Liang}},\ }\href@noop {} {\bibfield
  {journal} {\bibinfo  {journal} {Physical Review A}\ }\textbf {\bibinfo
  {volume} {91}},\ \bibinfo {pages} {032108} (\bibinfo {year}
  {2015})}\BibitemShut {NoStop}%
\bibitem [{\citenamefont {Home}\ \emph {et~al.}(2015)\citenamefont {Home},
  \citenamefont {Saha},\ and\ \citenamefont {Das}}]{Home2015}%
  \BibitemOpen
  \bibfield  {author} {\bibinfo {author} {\bibfnamefont {D.}~\bibnamefont
  {Home}}, \bibinfo {author} {\bibfnamefont {D.}~\bibnamefont {Saha}},\ and\
  \bibinfo {author} {\bibfnamefont {S.}~\bibnamefont {Das}},\ }\href@noop {}
  {\bibfield  {journal} {\bibinfo  {journal} {Physical Review A}\ }\textbf
  {\bibinfo {volume} {91}},\ \bibinfo {pages} {012102} (\bibinfo {year}
  {2015})}\BibitemShut {NoStop}%
\bibitem [{\citenamefont {Garuccio}(1995)}]{Garuccio1995}%
  \BibitemOpen
  \bibfield  {author} {\bibinfo {author} {\bibfnamefont {A.}~\bibnamefont
  {Garuccio}},\ }\href@noop {} {\bibfield  {journal} {\bibinfo  {journal}
  {Physical Review A}\ }\textbf {\bibinfo {volume} {52}},\ \bibinfo {pages}
  {2535} (\bibinfo {year} {1995})}\BibitemShut {NoStop}%
\bibitem [{\citenamefont {Van~Dam}\ \emph {et~al.}(2005)\citenamefont
  {Van~Dam}, \citenamefont {Gill},\ and\ \citenamefont
  {Grunwald}}]{vanDam2005}%
  \BibitemOpen
  \bibfield  {author} {\bibinfo {author} {\bibfnamefont {W.}~\bibnamefont
  {Van~Dam}}, \bibinfo {author} {\bibfnamefont {R.~D.}\ \bibnamefont {Gill}},\
  and\ \bibinfo {author} {\bibfnamefont {P.~D.}\ \bibnamefont {Grunwald}},\
  }\href@noop {} {\bibfield  {journal} {\bibinfo  {journal} {IEEE transactions
  on information theory}\ }\textbf {\bibinfo {volume} {51}},\ \bibinfo {pages}
  {2812} (\bibinfo {year} {2005})}\BibitemShut {NoStop}%
\bibitem [{\citenamefont {Ghirardi}\ and\ \citenamefont
  {Marinatto}(2008)}]{Ghirardi2008}%
  \BibitemOpen
  \bibfield  {author} {\bibinfo {author} {\bibfnamefont {G.}~\bibnamefont
  {Ghirardi}}\ and\ \bibinfo {author} {\bibfnamefont {L.}~\bibnamefont
  {Marinatto}},\ }\href@noop {} {\bibfield  {journal} {\bibinfo  {journal}
  {Physics Letters A}\ }\textbf {\bibinfo {volume} {372}},\ \bibinfo {pages}
  {1982} (\bibinfo {year} {2008})}\BibitemShut {NoStop}%
\bibitem [{\citenamefont {Braun}\ and\ \citenamefont {Choi}(2008)}]{Braun2008}%
  \BibitemOpen
  \bibfield  {author} {\bibinfo {author} {\bibfnamefont {D.}~\bibnamefont
  {Braun}}\ and\ \bibinfo {author} {\bibfnamefont {M.-S.}\ \bibnamefont
  {Choi}},\ }\href@noop {} {\bibfield  {journal} {\bibinfo  {journal} {Physical
  Review A}\ }\textbf {\bibinfo {volume} {78}},\ \bibinfo {pages} {032114}
  (\bibinfo {year} {2008})}\BibitemShut {NoStop}%
\bibitem [{\citenamefont {Man{\v{c}}inska}\ and\ \citenamefont
  {Wehner}(2014)}]{Mancinska2014}%
  \BibitemOpen
  \bibfield  {author} {\bibinfo {author} {\bibfnamefont {L.}~\bibnamefont
  {Man{\v{c}}inska}}\ and\ \bibinfo {author} {\bibfnamefont {S.}~\bibnamefont
  {Wehner}},\ }\href@noop {} {\bibfield  {journal} {\bibinfo  {journal}
  {Journal of Physics A: Mathematical and Theoretical}\ }\textbf {\bibinfo
  {volume} {47}},\ \bibinfo {pages} {424027} (\bibinfo {year}
  {2014})}\BibitemShut {NoStop}%
\bibitem [{\citenamefont {Dong}\ \emph {et~al.}(2020)\citenamefont {Dong},
  \citenamefont {Yang},\ and\ \citenamefont {Cao}}]{Dong2020}%
  \BibitemOpen
  \bibfield  {author} {\bibinfo {author} {\bibfnamefont {Z.}~\bibnamefont
  {Dong}}, \bibinfo {author} {\bibfnamefont {Y.}~\bibnamefont {Yang}},\ and\
  \bibinfo {author} {\bibfnamefont {H.}~\bibnamefont {Cao}},\ }\href@noop {}
  {\bibfield  {journal} {\bibinfo  {journal} {International Journal of
  Theoretical Physics}\ }\textbf {\bibinfo {volume} {59}},\ \bibinfo {pages}
  {1644} (\bibinfo {year} {2020})}\BibitemShut {NoStop}%
\bibitem [{\citenamefont {Boschi}\ \emph {et~al.}(1997)\citenamefont {Boschi},
  \citenamefont {Branca}, \citenamefont {De~Martini},\ and\ \citenamefont
  {Hardy}}]{boschi1997}%
  \BibitemOpen
  \bibfield  {author} {\bibinfo {author} {\bibfnamefont {D.}~\bibnamefont
  {Boschi}}, \bibinfo {author} {\bibfnamefont {S.}~\bibnamefont {Branca}},
  \bibinfo {author} {\bibfnamefont {F.}~\bibnamefont {De~Martini}},\ and\
  \bibinfo {author} {\bibfnamefont {L.}~\bibnamefont {Hardy}},\ }\href@noop {}
  {\bibfield  {journal} {\bibinfo  {journal} {Physical review letters}\
  }\textbf {\bibinfo {volume} {79}},\ \bibinfo {pages} {2755} (\bibinfo {year}
  {1997})}\BibitemShut {NoStop}%
\bibitem [{\citenamefont {Barbieri}\ \emph {et~al.}(2005)\citenamefont
  {Barbieri}, \citenamefont {De~Martini}, \citenamefont {Di~Nepi},\ and\
  \citenamefont {Mataloni}}]{Barbieri2005}%
  \BibitemOpen
  \bibfield  {author} {\bibinfo {author} {\bibfnamefont {M.}~\bibnamefont
  {Barbieri}}, \bibinfo {author} {\bibfnamefont {F.}~\bibnamefont
  {De~Martini}}, \bibinfo {author} {\bibfnamefont {G.}~\bibnamefont
  {Di~Nepi}},\ and\ \bibinfo {author} {\bibfnamefont {P.}~\bibnamefont
  {Mataloni}},\ }\href@noop {} {\bibfield  {journal} {\bibinfo  {journal}
  {Physics Letters A}\ }\textbf {\bibinfo {volume} {334}},\ \bibinfo {pages}
  {23} (\bibinfo {year} {2005})}\BibitemShut {NoStop}%
\bibitem [{\citenamefont {Sohbi}\ and\ \citenamefont {Kim}(2019)}]{Sohbi2019}%
  \BibitemOpen
  \bibfield  {author} {\bibinfo {author} {\bibfnamefont {A.}~\bibnamefont
  {Sohbi}}\ and\ \bibinfo {author} {\bibfnamefont {J.}~\bibnamefont {Kim}},\
  }\href@noop {} {\bibfield  {journal} {\bibinfo  {journal} {Physical Review
  A}\ }\textbf {\bibinfo {volume} {100}},\ \bibinfo {pages} {022117} (\bibinfo
  {year} {2019})}\BibitemShut {NoStop}%
\bibitem [{\citenamefont {Svozil}(2021)}]{Svozil2021}%
  \BibitemOpen
  \bibfield  {author} {\bibinfo {author} {\bibfnamefont {K.}~\bibnamefont
  {Svozil}},\ }\href@noop {} {\bibfield  {journal} {\bibinfo  {journal}
  {Physical Review A}\ }\textbf {\bibinfo {volume} {103}},\ \bibinfo {pages}
  {022204} (\bibinfo {year} {2021})}\BibitemShut {NoStop}%
\bibitem [{\citenamefont {Chen}\ \emph {et~al.}(2013)\citenamefont {Chen},
  \citenamefont {Cabello}, \citenamefont {Xu}, \citenamefont {Su},
  \citenamefont {Wu},\ and\ \citenamefont {Kwek}}]{chen2013}%
  \BibitemOpen
  \bibfield  {author} {\bibinfo {author} {\bibfnamefont {J.-L.}\ \bibnamefont
  {Chen}}, \bibinfo {author} {\bibfnamefont {A.}~\bibnamefont {Cabello}},
  \bibinfo {author} {\bibfnamefont {Z.-P.}\ \bibnamefont {Xu}}, \bibinfo
  {author} {\bibfnamefont {H.-Y.}\ \bibnamefont {Su}}, \bibinfo {author}
  {\bibfnamefont {C.}~\bibnamefont {Wu}},\ and\ \bibinfo {author}
  {\bibfnamefont {L.~C.}\ \bibnamefont {Kwek}},\ }\href@noop {} {\bibfield
  {journal} {\bibinfo  {journal} {Physical Review A}\ }\textbf {\bibinfo
  {volume} {88}},\ \bibinfo {pages} {062116} (\bibinfo {year}
  {2013})}\BibitemShut {NoStop}%
\bibitem [{\citenamefont {Chen}\ \emph {et~al.}(2017)\citenamefont {Chen},
  \citenamefont {Zhang}, \citenamefont {Wu}, \citenamefont {Wang},
  \citenamefont {Fickler},\ and\ \citenamefont {Karimi}}]{Chen2017}%
  \BibitemOpen
  \bibfield  {author} {\bibinfo {author} {\bibfnamefont {L.}~\bibnamefont
  {Chen}}, \bibinfo {author} {\bibfnamefont {W.}~\bibnamefont {Zhang}},
  \bibinfo {author} {\bibfnamefont {Z.}~\bibnamefont {Wu}}, \bibinfo {author}
  {\bibfnamefont {J.}~\bibnamefont {Wang}}, \bibinfo {author} {\bibfnamefont
  {R.}~\bibnamefont {Fickler}},\ and\ \bibinfo {author} {\bibfnamefont
  {E.}~\bibnamefont {Karimi}},\ }\href@noop {} {\bibfield  {journal} {\bibinfo
  {journal} {Physical Review A}\ }\textbf {\bibinfo {volume} {96}},\ \bibinfo
  {pages} {022115} (\bibinfo {year} {2017})}\BibitemShut {NoStop}%
\bibitem [{\citenamefont {Meng}\ \emph {et~al.}(2018)\citenamefont {Meng},
  \citenamefont {Zhou}, \citenamefont {Xu}, \citenamefont {Su}, \citenamefont
  {Gao}, \citenamefont {Yan},\ and\ \citenamefont {Chen}}]{Meng2018}%
  \BibitemOpen
  \bibfield  {author} {\bibinfo {author} {\bibfnamefont {H.-X.}\ \bibnamefont
  {Meng}}, \bibinfo {author} {\bibfnamefont {J.}~\bibnamefont {Zhou}}, \bibinfo
  {author} {\bibfnamefont {Z.-P.}\ \bibnamefont {Xu}}, \bibinfo {author}
  {\bibfnamefont {H.-Y.}\ \bibnamefont {Su}}, \bibinfo {author} {\bibfnamefont
  {T.}~\bibnamefont {Gao}}, \bibinfo {author} {\bibfnamefont {F.-L.}\
  \bibnamefont {Yan}},\ and\ \bibinfo {author} {\bibfnamefont {J.-L.}\
  \bibnamefont {Chen}},\ }\href@noop {} {\bibfield  {journal} {\bibinfo
  {journal} {Physical Review A}\ }\textbf {\bibinfo {volume} {98}},\ \bibinfo
  {pages} {062103} (\bibinfo {year} {2018})}\BibitemShut {NoStop}%
\bibitem [{\citenamefont {Rabelo}\ \emph {et~al.}(2012)\citenamefont {Rabelo},
  \citenamefont {Zhi},\ and\ \citenamefont {Scarani}}]{Rabelo2012}%
  \BibitemOpen
  \bibfield  {author} {\bibinfo {author} {\bibfnamefont {R.}~\bibnamefont
  {Rabelo}}, \bibinfo {author} {\bibfnamefont {L.~Y.}\ \bibnamefont {Zhi}},\
  and\ \bibinfo {author} {\bibfnamefont {V.}~\bibnamefont {Scarani}},\
  }\href@noop {} {\bibfield  {journal} {\bibinfo  {journal} {Physical Review
  Letters}\ }\textbf {\bibinfo {volume} {109}},\ \bibinfo {pages} {180401}
  (\bibinfo {year} {2012})}\BibitemShut {NoStop}%
\bibitem [{\citenamefont {Ac{\'\i}n}\ \emph {et~al.}(2005)\citenamefont
  {Ac{\'\i}n}, \citenamefont {Gill},\ and\ \citenamefont {Gisin}}]{Acin2005}%
  \BibitemOpen
  \bibfield  {author} {\bibinfo {author} {\bibfnamefont {A.}~\bibnamefont
  {Ac{\'\i}n}}, \bibinfo {author} {\bibfnamefont {R.}~\bibnamefont {Gill}},\
  and\ \bibinfo {author} {\bibfnamefont {N.}~\bibnamefont {Gisin}},\
  }\href@noop {} {\bibfield  {journal} {\bibinfo  {journal} {Physical review
  letters}\ }\textbf {\bibinfo {volume} {95}},\ \bibinfo {pages} {210402}
  (\bibinfo {year} {2005})}\BibitemShut {NoStop}%
\bibitem [{\citenamefont {Brunner}\ \emph {et~al.}(2005)\citenamefont
  {Brunner}, \citenamefont {Gisin},\ and\ \citenamefont
  {Scarani}}]{Brunner2005}%
  \BibitemOpen
  \bibfield  {author} {\bibinfo {author} {\bibfnamefont {N.}~\bibnamefont
  {Brunner}}, \bibinfo {author} {\bibfnamefont {N.}~\bibnamefont {Gisin}},\
  and\ \bibinfo {author} {\bibfnamefont {V.}~\bibnamefont {Scarani}},\
  }\href@noop {} {\bibfield  {journal} {\bibinfo  {journal} {New Journal of
  Physics}\ }\textbf {\bibinfo {volume} {7}},\ \bibinfo {pages} {88} (\bibinfo
  {year} {2005})}\BibitemShut {NoStop}%
\bibitem [{\citenamefont {Junge}\ and\ \citenamefont
  {Palazuelos}(2011)}]{Junge2011}%
  \BibitemOpen
  \bibfield  {author} {\bibinfo {author} {\bibfnamefont {M.}~\bibnamefont
  {Junge}}\ and\ \bibinfo {author} {\bibfnamefont {C.}~\bibnamefont
  {Palazuelos}},\ }\href@noop {} {\bibfield  {journal} {\bibinfo  {journal}
  {Communications in Mathematical Physics}\ }\textbf {\bibinfo {volume}
  {306}},\ \bibinfo {pages} {695} (\bibinfo {year} {2011})}\BibitemShut
  {NoStop}%
\bibitem [{\citenamefont {Liang}\ \emph {et~al.}(2011)\citenamefont {Liang},
  \citenamefont {V{\'e}rtesi},\ and\ \citenamefont {Brunner}}]{Liang2011}%
  \BibitemOpen
  \bibfield  {author} {\bibinfo {author} {\bibfnamefont {Y.-C.}\ \bibnamefont
  {Liang}}, \bibinfo {author} {\bibfnamefont {T.}~\bibnamefont {V{\'e}rtesi}},\
  and\ \bibinfo {author} {\bibfnamefont {N.}~\bibnamefont {Brunner}},\
  }\href@noop {} {\bibfield  {journal} {\bibinfo  {journal} {Physical Review
  A}\ }\textbf {\bibinfo {volume} {83}},\ \bibinfo {pages} {022108} (\bibinfo
  {year} {2011})}\BibitemShut {NoStop}%
\bibitem [{\citenamefont {Vidick}\ and\ \citenamefont
  {Wehner}(2011)}]{Vidick2011}%
  \BibitemOpen
  \bibfield  {author} {\bibinfo {author} {\bibfnamefont {T.}~\bibnamefont
  {Vidick}}\ and\ \bibinfo {author} {\bibfnamefont {S.}~\bibnamefont
  {Wehner}},\ }\href@noop {} {\bibfield  {journal} {\bibinfo  {journal}
  {Physical Review A}\ }\textbf {\bibinfo {volume} {83}},\ \bibinfo {pages}
  {052310} (\bibinfo {year} {2011})}\BibitemShut {NoStop}%
\bibitem [{\citenamefont {Dilley}\ and\ \citenamefont
  {Chitambar}(2018)}]{Diley2018}%
  \BibitemOpen
  \bibfield  {author} {\bibinfo {author} {\bibfnamefont {D.}~\bibnamefont
  {Dilley}}\ and\ \bibinfo {author} {\bibfnamefont {E.}~\bibnamefont
  {Chitambar}},\ }\href@noop {} {\bibfield  {journal} {\bibinfo  {journal}
  {Physical Review A}\ }\textbf {\bibinfo {volume} {97}},\ \bibinfo {pages}
  {062313} (\bibinfo {year} {2018})}\BibitemShut {NoStop}%
\bibitem [{\citenamefont {Irvine}\ \emph {et~al.}(2005)\citenamefont {Irvine},
  \citenamefont {Hodelin}, \citenamefont {Simon},\ and\ \citenamefont
  {Bouwmeester}}]{irvine2005}%
  \BibitemOpen
  \bibfield  {author} {\bibinfo {author} {\bibfnamefont {W.~T.}\ \bibnamefont
  {Irvine}}, \bibinfo {author} {\bibfnamefont {J.~F.}\ \bibnamefont {Hodelin}},
  \bibinfo {author} {\bibfnamefont {C.}~\bibnamefont {Simon}},\ and\ \bibinfo
  {author} {\bibfnamefont {D.}~\bibnamefont {Bouwmeester}},\ }\href@noop {}
  {\bibfield  {journal} {\bibinfo  {journal} {Physical review letters}\
  }\textbf {\bibinfo {volume} {95}},\ \bibinfo {pages} {030401} (\bibinfo
  {year} {2005})}\BibitemShut {NoStop}%
\bibitem [{\citenamefont {Lundeen}\ and\ \citenamefont
  {Steinberg}(2009)}]{Lundeen2009}%
  \BibitemOpen
  \bibfield  {author} {\bibinfo {author} {\bibfnamefont {J.~S.}\ \bibnamefont
  {Lundeen}}\ and\ \bibinfo {author} {\bibfnamefont {A.~M.}\ \bibnamefont
  {Steinberg}},\ }\href@noop {} {\bibfield  {journal} {\bibinfo  {journal}
  {Physical review letters}\ }\textbf {\bibinfo {volume} {102}},\ \bibinfo
  {pages} {020404} (\bibinfo {year} {2009})}\BibitemShut {NoStop}%
\bibitem [{\citenamefont {Yokota}\ \emph {et~al.}(2009)\citenamefont {Yokota},
  \citenamefont {Yamamoto}, \citenamefont {Koashi},\ and\ \citenamefont
  {Imoto}}]{Yokota2009}%
  \BibitemOpen
  \bibfield  {author} {\bibinfo {author} {\bibfnamefont {K.}~\bibnamefont
  {Yokota}}, \bibinfo {author} {\bibfnamefont {T.}~\bibnamefont {Yamamoto}},
  \bibinfo {author} {\bibfnamefont {M.}~\bibnamefont {Koashi}},\ and\ \bibinfo
  {author} {\bibfnamefont {N.}~\bibnamefont {Imoto}},\ }\href@noop {}
  {\bibfield  {journal} {\bibinfo  {journal} {New Journal of Physics}\ }\textbf
  {\bibinfo {volume} {11}},\ \bibinfo {pages} {033011} (\bibinfo {year}
  {2009})}\BibitemShut {NoStop}%
\bibitem [{\citenamefont {Luo}\ \emph {et~al.}(2018)\citenamefont {Luo},
  \citenamefont {Su}, \citenamefont {Huang}, \citenamefont {Wang},
  \citenamefont {Yang}, \citenamefont {Li}, \citenamefont {Liu}, \citenamefont
  {Chen}, \citenamefont {Lu},\ and\ \citenamefont {Pan}}]{luo2018}%
  \BibitemOpen
  \bibfield  {author} {\bibinfo {author} {\bibfnamefont {Y.-H.}\ \bibnamefont
  {Luo}}, \bibinfo {author} {\bibfnamefont {H.-Y.}\ \bibnamefont {Su}},
  \bibinfo {author} {\bibfnamefont {H.-L.}\ \bibnamefont {Huang}}, \bibinfo
  {author} {\bibfnamefont {X.-L.}\ \bibnamefont {Wang}}, \bibinfo {author}
  {\bibfnamefont {T.}~\bibnamefont {Yang}}, \bibinfo {author} {\bibfnamefont
  {L.}~\bibnamefont {Li}}, \bibinfo {author} {\bibfnamefont {N.-L.}\
  \bibnamefont {Liu}}, \bibinfo {author} {\bibfnamefont {J.-L.}\ \bibnamefont
  {Chen}}, \bibinfo {author} {\bibfnamefont {C.-Y.}\ \bibnamefont {Lu}},\ and\
  \bibinfo {author} {\bibfnamefont {J.-W.}\ \bibnamefont {Pan}},\ }\href@noop
  {} {\bibfield  {journal} {\bibinfo  {journal} {Science bulletin}\ }\textbf
  {\bibinfo {volume} {63}},\ \bibinfo {pages} {1611} (\bibinfo {year}
  {2018})}\BibitemShut {NoStop}%
\bibitem [{\citenamefont {Matsukevich}\ \emph {et~al.}(2008)\citenamefont
  {Matsukevich}, \citenamefont {Maunz}, \citenamefont {Moehring}, \citenamefont
  {Olmschenk},\ and\ \citenamefont {Monroe}}]{matsukevich2008}%
  \BibitemOpen
  \bibfield  {author} {\bibinfo {author} {\bibfnamefont {D.}~\bibnamefont
  {Matsukevich}}, \bibinfo {author} {\bibfnamefont {P.}~\bibnamefont {Maunz}},
  \bibinfo {author} {\bibfnamefont {D.~L.}\ \bibnamefont {Moehring}}, \bibinfo
  {author} {\bibfnamefont {S.}~\bibnamefont {Olmschenk}},\ and\ \bibinfo
  {author} {\bibfnamefont {C.}~\bibnamefont {Monroe}},\ }\href@noop {}
  {\bibfield  {journal} {\bibinfo  {journal} {Physical Review Letters}\
  }\textbf {\bibinfo {volume} {100}},\ \bibinfo {pages} {150404} (\bibinfo
  {year} {2008})}\BibitemShut {NoStop}%
\bibitem [{\citenamefont {Hofmann}\ \emph {et~al.}(2012)\citenamefont
  {Hofmann}, \citenamefont {Krug}, \citenamefont {Ortegel}, \citenamefont
  {G{\'e}rard}, \citenamefont {Weber}, \citenamefont {Rosenfeld},\ and\
  \citenamefont {Weinfurter}}]{hofmann2012}%
  \BibitemOpen
  \bibfield  {author} {\bibinfo {author} {\bibfnamefont {J.}~\bibnamefont
  {Hofmann}}, \bibinfo {author} {\bibfnamefont {M.}~\bibnamefont {Krug}},
  \bibinfo {author} {\bibfnamefont {N.}~\bibnamefont {Ortegel}}, \bibinfo
  {author} {\bibfnamefont {L.}~\bibnamefont {G{\'e}rard}}, \bibinfo {author}
  {\bibfnamefont {M.}~\bibnamefont {Weber}}, \bibinfo {author} {\bibfnamefont
  {W.}~\bibnamefont {Rosenfeld}},\ and\ \bibinfo {author} {\bibfnamefont
  {H.}~\bibnamefont {Weinfurter}},\ }\href@noop {} {\bibfield  {journal}
  {\bibinfo  {journal} {Science}\ }\textbf {\bibinfo {volume} {337}},\ \bibinfo
  {pages} {72} (\bibinfo {year} {2012})}\BibitemShut {NoStop}%
\bibitem [{\citenamefont {Das}\ and\ \citenamefont {Paul}(2020)}]{das2020}%
  \BibitemOpen
  \bibfield  {author} {\bibinfo {author} {\bibfnamefont {S.}~\bibnamefont
  {Das}}\ and\ \bibinfo {author} {\bibfnamefont {G.}~\bibnamefont {Paul}},\
  }\href@noop {} {\bibfield  {journal} {\bibinfo  {journal} {ACM Transactions
  on Quantum Computing}\ }\textbf {\bibinfo {volume} {1}},\ \bibinfo {pages}
  {1} (\bibinfo {year} {2020})}\BibitemShut {NoStop}%
\bibitem [{\citenamefont {Hou}\ \emph {et~al.}(2021)\citenamefont {Hou},
  \citenamefont {Ding}, \citenamefont {Wang}, \citenamefont {Zhang},\ and\
  \citenamefont {He}}]{Hou2021}%
  \BibitemOpen
  \bibfield  {author} {\bibinfo {author} {\bibfnamefont {T.}~\bibnamefont
  {Hou}}, \bibinfo {author} {\bibfnamefont {D.}~\bibnamefont {Ding}}, \bibinfo
  {author} {\bibfnamefont {C.}~\bibnamefont {Wang}}, \bibinfo {author}
  {\bibfnamefont {X.-c.}\ \bibnamefont {Zhang}},\ and\ \bibinfo {author}
  {\bibfnamefont {Y.-q.}\ \bibnamefont {He}},\ }\href@noop {} {\bibfield
  {journal} {\bibinfo  {journal} {International Journal of Theoretical
  Physics}\ }\textbf {\bibinfo {volume} {60}},\ \bibinfo {pages} {1972}
  (\bibinfo {year} {2021})}\BibitemShut {NoStop}%
\bibitem [{exc()}]{excludestate}%
  \BibitemOpen
  \href@noop {} {}\bibinfo {note} {Eq. (8) and (9) cannot describe an abitrary
  state. For example, state $|X\rangle = 1|100\rangle + 2|010\rangle +
  3|001\rangle + 4|011\rangle$ up to some normalization. Following Eq. (9), it
  would be written as $|X\rangle = A_1B_2B_3|u_1v_2v_3\rangle +
  B_1A_2B_3|v_1u_2v_3\rangle + B_1B_2A_3|v_1v_2u_3\rangle +
  B_1A_2A_3|v_1u_2u_3\rangle$. We could not find the appropriate $A_k$, $B_k$
  coefficients for such state.}\BibitemShut {Stop}%
\bibitem [{\citenamefont {Aharonov}\ \emph {et~al.}(2002)\citenamefont
  {Aharonov}, \citenamefont {Botero}, \citenamefont {Popescu}, \citenamefont
  {Reznik},\ and\ \citenamefont {Tollaksen}}]{Aharonov2002}%
  \BibitemOpen
  \bibfield  {author} {\bibinfo {author} {\bibfnamefont {Y.}~\bibnamefont
  {Aharonov}}, \bibinfo {author} {\bibfnamefont {A.}~\bibnamefont {Botero}},
  \bibinfo {author} {\bibfnamefont {S.}~\bibnamefont {Popescu}}, \bibinfo
  {author} {\bibfnamefont {B.}~\bibnamefont {Reznik}},\ and\ \bibinfo {author}
  {\bibfnamefont {J.}~\bibnamefont {Tollaksen}},\ }\href@noop {} {\bibfield
  {journal} {\bibinfo  {journal} {Physics Letters A}\ }\textbf {\bibinfo
  {volume} {301}},\ \bibinfo {pages} {130} (\bibinfo {year}
  {2002})}\BibitemShut {NoStop}%
\bibitem [{git()}]{github}%
  \BibitemOpen
  \href@noop {} {}\bibinfo {note} {The code in this work can be accessed at:
  https://github.com/mx73/Testing-QM-on-NISQ.}\BibitemShut {Stop}%
\bibitem [{\citenamefont {Tran}\ \emph {et~al.}(2022)\citenamefont {Tran},
  \citenamefont {Nguyen}, \citenamefont {Le},\ and\ \citenamefont
  {Nguyen}}]{Tran2022}%
  \BibitemOpen
  \bibfield  {author} {\bibinfo {author} {\bibfnamefont {D.~M.}\ \bibnamefont
  {Tran}}, \bibinfo {author} {\bibfnamefont {D.~V.}\ \bibnamefont {Nguyen}},
  \bibinfo {author} {\bibfnamefont {B.~H.}\ \bibnamefont {Le}},\ and\ \bibinfo
  {author} {\bibfnamefont {H.~Q.}\ \bibnamefont {Nguyen}},\ }\href@noop {}
  {\bibfield  {journal} {\bibinfo  {journal} {EPJ Quantum Technology}\ }\textbf
  {\bibinfo {volume} {9}},\ \bibinfo {pages} {6} (\bibinfo {year}
  {2022})}\BibitemShut {NoStop}%
\end{thebibliography}%

\newpage
\appendix
\sloppy

\section{Some detail calculations}
Here, we derive some detail calculations in the main text. 

\emph{Transformation between bases}

In $\{|u\ra_k, |v\ra_k\}^{\otimes n}$, a general state writes
\begin{align}\label{eq:app:state}
	|\Phi\ra = a_1|u_1u_2\cdots u_n\ra + a_2|u_1u_2\cdots v_n\ra + \cdots + a_{2^n}|v_1v_2\cdots v_n\ra,
\end{align}
with complex $a_k$ satisfies the completeness relation $\sum_{k = 1}^{2^n}|a_k|^2 = 1$. To transfer this state into the $\{|c\ra_k, |d\ra_k\}^{\otimes n}$ basis, we start from the state $|c_1c_2\cdots c_n\ra$ and use Eq.~\ref{eq:bs2} to obtain
\begin{align}\label{eq:app:a1}
	\notag|c_1c_2\cdots c_n\ra &= \bigl(A_1|u_1\ra+B_1|v_1\ra\bigr) \otimes	\bigl(A_2|u_2\ra+B_2|v_2\ra\bigr) \otimes \cdots \otimes  \bigl(A_n|u_n\ra+B_n|v_n\ra\bigr)\\
	&= A_1A_2\cdots A_n |u_1u_2\cdots u_n\ra + A_1A_2\cdots B_n |u_1u_2\cdots v_n\ra + \cdots + B_1B_2\cdots B_n |v_1v_2\cdots v_n\ra.
\end{align}
Without loss of generality, in Eq.~\ref{eq:app:state} we choose $a_1 = A_1A_2\cdots A_n, a_2 = A_1A_2\cdots B_n$ and so on, then $|\Phi\rangle = |c_1c_2\cdots c_n\ra.$ The state of interested in Eq.~\eqref{eq:psi1.1} can be rewrite as
\begin{align}
	\notag	|\Psi_n\ra &= N\Bigl[A_1A_2\cdots B_n |u_1u_2\cdots v_n\ra + A_1A_2\cdots B_{n-1}A_n |u_1u_2\cdots v_{n-1}u_n\ra + \cdots \notag \\ 
	& \hspace{2cm}+ B_1B_2\cdots B_n|v_1v_2\cdots v_n\ra\Bigr] \notag\\
	&= N\Bigl[\mathcal{A}_{\Omega\backslash \{n\}}B_n |u_1u_2\cdots v_n\ra + \mathcal{A}_{\Omega\backslash \{n-1\}} B_{n-1}	|u_1u_2\cdots v_{n-1}u_n\ra + \cdots + \mathcal{B}_{\Omega }|v_1v_2\cdots v_n\ra\Bigr] \notag \\ 
	&=N\sum_{\alpha \subseteq \mathcal{P}(\Omega)\setminus \{\Omega\}} \mathcal{A}_\alpha \mathcal{B}_{\overline{\alpha}} \bigotimes_{i\in \alpha} |u_i\ra \otimes \bigotimes_{j\in \overline{\alpha}} |v_j\ra. \label{apd:psi1.3}
\end{align}

\emph{Normalization constant}

The normalization constant for our state is calculated as
\begin{align}\label{eq:app:state:nor}
	\notag \la\Psi_n|\Psi_n\ra &= N^2 \Bigl[A_1^*A_2^*\cdots B_n^*\la u_1u_2\cdots v_n| + \cdots + 
	B_1^*B_2^*\cdots B_n^*\la v_1v_2\cdots v_n|\Bigr]  \\
	\notag &\hspace{4cm} \times \Bigl[A_1A_2\cdots B_n |u_1u_2\cdots v_n\ra + \cdots + B_1B_2\cdots B_n|v_1v_2\cdots v_n\ra\Bigr]\\
	\notag &= N^2 \Bigl[|A_1|^2|A_2|^2\cdots |B_n|^2 + \cdots + |B_1|^2|B_2|^2\cdots |B_n|^2\Bigr]\\
	\notag &= N^2 \Bigl[1-|A_1|^2|A_2|^2\cdots |A_n|^2 \Bigr]\\	               
	&=N^2 \Bigl[1-|\mathcal{A}_\Omega |^2\Bigr],
\end{align}
where $|\mathcal{A}_\Omega |^2 = \Pi_{k\in\Omega }|A_k|^2$. Then, from the normalization condition $\la\Psi_n|\Psi_n\ra = 1$, we obtain $N = \dfrac{1}{\sqrt{1-|\mathcal{A}_\Omega |^2}}$.

\emph{Success probability}

The success probability $P({\mathcal D}_\alpha)$ is calculated by first expanding  Eq.~\eqref{eq:psi1.1} into the $\{|c\ra, |d\ra\}^{\otimes n}$ basis as 
\begin{align}
	\notag|\Psi_n\ra &= N\Bigl[|c_1 \cdots c_n\ra - \mathcal{A}_\Omega \bigotimes_{k=1}^{n} \Bigl(A^*_k|c_k\ra-B_k|d_k\ra\Bigr)\Bigr]\\
	\notag &= N\Bigl[|c_1 \cdots c_n\ra - \mathcal{A}_\Omega \Bigl(A^*_1|c_1\ra-B_1|d_1\ra\Bigr)\\
	& \notag \hspace{20pt} \otimes \Bigl(A^*_2|c_2\ra-B_2|d_2\ra\Bigr) \otimes \cdots \otimes \Bigl(A^*_n|c_n\ra-B_n|d_n\ra\Bigr) \Bigr]\\
	& \notag = N\Bigl[ \bigl(1-|\mathcal{A}_\Omega |^2\bigr) |c_1 \cdots c_n\ra + \mathcal{A}_\Omega \sum_{k=1}^n \dfrac{\mathcal{A}^*_\Omega} {A^*_k}B_k |c_1 \cdots d_k \cdots c_n\ra \\
	& \hspace{20pt} + \mathcal{A}_\Omega \sum_{k=1}^{n}\sum_{l=k+1}^{n} \dfrac{\mathcal{A}^*_\Omega } {A^*_kA^*_l}B_kB_l |c_1 \cdots d_k \cdots d_l \cdots c_n\ra + \cdots \Bigr].
\end{align}
There are many substates containing two or more qubits with $|\cdots d_k\cdots d_l\cdots\ra$. They all satisfy Eq.\eqref{eq:psicon3}. From the binomial theorem, there are $\sum_{k=2}^{n} \binom{n}{k} = 2^n-n-1$ of them, and their combined probability is the complementary probability of substates containing one or less  $|\cdots d_k \dots\ra$. To find their success probabilities, we focus on the first two terms: the term without $|d\ra$ and the term with one $|d\ra$). Let their probabilities be $p_0$ and $p_1$, respectively
\begin{align}
	p_0 = N^2 (1-|\mathcal{A}_\Omega |^2)^2 \ \text { and }
	p_1 = N^2|\mathcal{A}_\Omega |^4 \sum_{k = 1}^n\dfrac{|B_k|^2}{|A_k|^2}.
\end{align}
Then, the success probability is $1- (p_0 + p_1)$, and
\begin{align}
	P_\text{success}& = 1 - (p_0 + p_1) \notag\\
	& = 1 - N^2 \Bigl[ \bigl(1-|\mathcal{A}_\Omega |^2\bigr)^2
	+|\mathcal{A}_\Omega |^4 \sum_{k = 1}^n\frac{|B_k|^2}{|A_k|^2} \Bigr] \notag \\
	&=|\mathcal{A}_\Omega|^2-\dfrac{|\mathcal{A}_\Omega|^4}{1-|\mathcal{A}_\Omega|^2}\sum_{k = 1}^n\frac{1-|A_k|^2}{|A_k|^2}.
\end{align}
It rewrites explicitly as
\begin{align}
	P_\text{success}
	& = \prod_{i=1}^{n}|A{_i}|^{2}-\frac{\prod_{i=1}^{n}|A{_i}|^{4}}{1-\prod_{i=1}^{n}|A{_i}|^{2}}\sum_{k=1}^{n}\frac{1-|A{_k}|^2}{|A{_k}|^2} \\
	& = \prod_{i=1}^{n}|A{_i}|^{2}+\frac{n\prod_{i=1}^{n}|A{_i}|^{4}}{1-\prod_{i=1}^{n}|A{_i}|^{2}}-\frac{\prod_{i=1}^{n}|A{_i}|^{4}}{1-\prod_{i=1}^{n}|A{_i}|^{2}}\sum_{k=1}^{n}\frac{1}{|A{_k}|^2}.
\end{align}
From the arithmetic mean - geometric mean inequality, $\sum_{k=1}^{n} \frac{1}{|A{_k}|^{2}} \ge n \sqrt[n]{\prod_{k=1}^{n} \frac{1}{|A{_k}|^{2}}}$, we have
\begin{equation}
	P_\text{success} \le \prod_{i=1}^{n}|A{_i}|^{2}+\frac{n\prod_{i=1}^{n}|A{_i}|^{4}}{1-\prod_{i=1}^{n}|A{_i}|^{2}}\left(1-\sqrt[n]{\frac{1}{\prod_{k=1}^{n} |A{_k}|^{2}}}\right).
\end{equation}
The equality holds when $|A_1| = |A_2| = ... = |A_n| = A$, in which case, $P_\text{success}$ reaches its maximal value:
\begin{equation}
	\begin{aligned}
	P_\text{success} &= A^{2n} + n\frac{A^{4n}}{1-A^{2n}}-n\frac{A^{4n-2}}{1-A^{2n}}.
	\end{aligned}
\end{equation}
Eq.~\eqref{eq:app:non1} becomes
\begin{align}
	P_\text{success}=A^{2n}-n\frac{A^{4n-2}(1-A^{2})}{1-A^{2n}}. 
\end{align}

\section{Creating $|\Psi_n\ra$: comparing post-selection scheme to the tradditional approach}
\label{PostSelection}
Connecting with previous literature \cite{wu1996, ghosh1998}, we construct $|\Psi_n \ra$ using their method. Let us consider a general 3-particle state \cite{wu1996} satisfying normalizing condition $|a|^2+|b|^2+|c|^2+|d|^2+|e|^2+|f|^2+|g|^2+|h|^2=1$, thus
\begin{equation}\label{eq:uni1}
	\begin{aligned}
		|\psi_3\rangle&= a\left|u_{1}\right\rangle\left|u_{2}\right\rangle\left|u_{3}\right\rangle+b\left|u_{1}\right\rangle\left|u_{2}\right\rangle\left|v_{3}\right\rangle+c\left|u_{1}\right\rangle\left|v_{2}\right\rangle\left|u_{3}\right\rangle+d\left|u_{1}\right\rangle\left|v_{2}\right\rangle\left|v_{3}\right\rangle \\
		&+e\left|v_{1}\right\rangle\left|u_{2}\right\rangle\left|u_{3}\right\rangle+f\left|v_{1}\right\rangle\left|u_{2}\right\rangle\left|v_{3}\right\rangle+g\left|v_{1}\right\rangle\left|v_{2}\right\rangle\left|u_{3}\right\rangle+h\left|v_{1}\right\rangle\left|v_{2}\right\rangle\left|v_{3}\right\rangle.
	\end{aligned}
\end{equation}
Let $a=0$ and $\frac{b}{f} = \frac{c}{g} = \frac{d}{h}, \frac{b}{d} = \frac{e}{g}$. Stem from the post-selection rule, the constraints constitute five degrees of freedom, leaving $8-5=3$ left from a general 3-particle state. For convenience, we expand the equalities to $\frac{b}{f} = \frac{c}{g} = \frac{d}{h}, \frac{b}{d} = \frac{e}{g} = \frac{f}{h}, \frac{c}{d} = \frac{e}{f} = \frac{g}{h}$. Then,
\begin{align}
|c_{1}\ra & = \frac{d|u_{1}\ra+h|v_{1}\ra}{(|d|^{2}+|h|^{2})^{1/2}}=\frac{c|u_{1}\ra+g|v_{1}\ra}{(|c|^{2}+|g|^{2})^{1/2}}=\frac{b|u_{1}\ra+f|v_{1}\ra}{(|b|^{2}+|f|^{2})^{1/2}},\notag\\
|c_{2}\ra & = \frac{f|u_{2}\ra+h|v_{2}\ra}{(|f|^{2}+|h|^{2})^{1/2}}=\frac{e|u_{2}\ra+g|v_{2}\ra}{(|e|^{2}+|g|^{2})^{1/2}}=\frac{b|u_{2}\ra+d|v_{2}\ra}{(|b|^{2}+|d|^{2})^{1/2}}, \notag\\
|c_{3}\ra & = \frac{g|u_{3}\ra+h|v_{3}\ra}{(|g|^{2}+|h|^{2})^{1/2}}=\frac{e|u_{3}\ra+f|v_{3}\ra}{(|e|^{2}+|f|^{2})^{1/2}}=\frac{c|u_{3}\ra+d|v_{3}\ra}{(|c|^{2}+|d|^{2})^{1/2}},\notag\\
|d_{1}\ra & = \frac{-h^{*}|u_{1}\ra+d^{*}|v_{1}\ra}{(|d|^{2}+|h|^{2})^{1 / 2}}=\frac{-g^{*}|u_{1}\ra+c^{*}|v_{1}\ra}{(|c|^{2}+|g|^{2})^{1 / 2}}=\frac{-f^{*}|u_{1}\ra+b^{*}|v_{1}\ra}{(|b|^{2}+|f|^{2})^{1/2}},\notag\\
|d_{2}\ra & = \frac{-h^{*}|u_{2}\ra+f^{*}|v_{2}\ra}{(|f|^{2}+|h|^{2})^{1 / 2}}=\frac{-g^{*}|u_{2}\ra+e^{*}|v_{2}\ra}{(|e|^{2}+|g|^{2})^{1 / 2}}=\frac{-d^{*}|u_{2}\ra+b^{*}|v_{2}\ra}{(|b|^{2}+|d|^{2})^{1/2}},\notag\\
|d_{3}\ra & = \frac{-h^{*}|u_{3}\ra+g^{*}|v_{3}\ra}{(|g|^{2}+|h|^{2})^{1 / 2}}=\frac{-f^{*}|u_{3}\ra+e^{*}|v_{3}\ra}{(|e|^{2}+|f|^{2})^{1 / 2}}=\frac{-d^{*}|u_{3}\ra+c^{*}|v_{3}\ra}{(|c|^{2}+|d|^{2})^{1/2}}.
\end{align}
Comparing $|\psi_3\ra$ in Eq.~\eqref{eq:uni1} to the 3-particle $|\Psi_3\ra$ in the main article,
\begin{align}
	|\Psi_3\ra &= N(A_{1} A_{2} B_{3} |u_{1}\ra |u_{2}\ra |v_{3}\ra 
	+ A_{1} B_{2} A_{3} |u_{1}\ra |v_{2}\ra |u_{3}\ra 
	+ A_{1} B_{2} B_{3} |u_{1}\ra |v_{2}\ra |v_{3}\ra\notag\\
	&+ B_{1} A_{2} A_{3} |v_{1}\ra |u_{2}\ra |u_{3}\ra 
	+ B_{1} A_{2} B_{3} |v_{1}\ra |u_{2}\ra  |v_{3}\ra 
	+ B_{1} B_{2} A_{3} |v_{1}\ra |v_{2}\ra |u_{3}\ra
	+ B_{1} B_{2} B_{3} |v_{1}\ra |v_{2}\ra |v_{3}\ra),
\end{align}
we see that they are the same. For example, $|c_{1}\ra=\frac{d|u_{1}\ra+h|v_{1}\ra}{(|d|^{2}+|h|^{2})^{1 / 2}} =\frac{A_{1} B_{2} B_{3}|u_{1}\ra+B_{1} B_{2} B_{3}|v_{1}\ra}{(|A_{1} B_{2} B_{3}|^{2}+|B_{1} B_{2} B_{3}|^{2})^{1 / 2}} = A_1|u_{1}\ra + B_1|v_{1}\ra$, knowing $|z_1z_2|^2=|z_1|^2|z_2|^2 \ \forall z_1, z_2 \in \mathbb{C}$. We analyze $|\Psi_3\ra$ in the mixed bases $\{|c_1\ra,|d_1\ra\}$, $\{|u_2\ra,|v_2\ra\}$, $\{|u_3\ra,|v_3\ra\}$,
\begin{equation}\label{eq:uni2}
	\begin{aligned}
	|\Psi_3\rangle&=(|d|^{2}+|h|^{2})^{1 / 2}|c_{1}\rangle|v_{2}\rangle|v_{3}\rangle
		+\frac{d^{*} b+h^{*} f}{(|d|^{2}+|h|^{2})^{1 / 2}}|c_{1}\rangle  |u_{2}\rangle|v_{3}\rangle \\
		&+\frac{d^{*} c+h^{*} g}{(|d|^{2}+|h|^{2})^{1 / 2}}|c_{1}\rangle |v_{2}\rangle|u_{3}\rangle
		+\frac{e^{*} h|c_{1}\rangle+d e|d_{1}\rangle}{(|d|^{2}+|h|^{2})^{1 / 2}}|u_{2}\rangle|u_{3}\rangle
\end{aligned}\end{equation}
From \eqref{eq:uni1} and  \eqref{eq:uni2} the first and second Hardy's conditions can be deduced, which are $P(U_1U_2U_3)=0$ and $P(U_2U_3|D_1) =1$. Two more second Hardy's conditions are found by choosing a different $\{|c_i\ra,|d_i|\ra\}$ basis, and the third conditions is derived by converting $|\Psi_3\ra$ fully into $\{|c\ra,|d|\ra\}$ bases.

\section{Success probabilities, von Neumann entropies and lowest-valued eigenvalues according to the Peres-Horodecki criterion of optimal cases}

In this section, we present extra data related to the arguments made in the article. Following previous notation on the Hilbert space $\mathcal{H} = \mathcal{H}_{\mathcal{A}} \otimes \mathcal{H}_{\mathcal{B}}$, we analyze a different bipartition where $\mathcal{A}$'s is that of one particle and $\mathcal{B}$ for the rest of the system, to introduce the von Neumann entropy at 1 vs. all bipartition $S(\rho_{1})=-\operatorname{tr}(\rho_{1} \log_2 \rho_{1})$. For the PPT criterion, we define $\rho^{T_B} = (I_A\otimes T_B)(\rho)$ as the partial transpose according to the aforementioned bipartition  with $I_A$ is the identity map in $H_A$ and $T_B$ is the transposition map in $H_B$. We then calculate PPT$(\rho_1) = \lambda_{min}$ as the minimal eigenvalue of $\rho^{T_B}$. Simulating $|\Psi_n\ra$, we find the optimal success probabilities from quantum simulation closely reflect theoretical results found by analytical formula. As increasing $n$, the data suggest a gradual increase in the success probability towards 0.152 (15.2\%), meanwhile, we also observe a gradual decrease in entanglement as $n$ increases. Nevertheless, calculations at large $n$ are difficult due to limitation in computation and memory resources.

\begin{table*}[h]
	\caption{\label{table1} Maximal success probabilities as found by Python (Py) optimization and Qiskit (Qi) simulations, von Neumann entropies at the 1 vs. all and half-chain bipartition, lowest-valued eigenvalues according to the Peres-Horodecki criterion at the 1 vs. all bipartition, and their associating $A$ values for optimal $|\Psi_n \rangle$, $n$ from 2 to 29.}
	\begin{ruledtabular}
		\begin{tabular}{c c c c c c c c c}
			n & 2 & 3 & 4 & 5 & 6 & 7 & 8 & 9 \\ 
			\hline\hline
			$P_\text{success}$(Py) & 0.09017 & 0.114233 & 0.125434 & 0.131916 & 0.136143 & 0.139117 & 0.141324 & 0.143026 \\
			$P_\text{success}$(Qi) & 0.090165 & 0.114268 & 0.125638 & 0.131913 & 0.136207 & 0.139345 & 0.141488 & 0.14307 \\ 
			$A$ & 0.786151& 0.858248& 0.893961& 0.915293& 0.929478& 0.939592 & 0.947169 & 0.953057 \\
			S$(\rho_1)$ & 0.674249& 0.631515& 0.56225 & 0.502161& 0.452987& 0.412712 & 0.379316 & 0.351233 \\
			S$(\rho_{{\lfloor n/2 \rfloor}})$ & 0.674249& 0.631515& 0.69472 & 0.677888& 0.70075 & 0.69193 & 0.703637 & 0.698231 \\
			PPT$(\rho_1)$ & -0.381966& -0.365522& -0.338231& -0.313817& -0.293234& -0.275904& -0.261163& -0.248473 \\
		\end{tabular}
	\end{ruledtabular}
	\begin{ruledtabular}
		\begin{tabular}{c c c c c c c c}
			n & 10 & 11 & 12 & 13 & 14 & 15 & 16 \\ 
			\hline\hline
			$P_\text{success}$(Py) &0.14438 & 0.145482 & 0.146396 & 0.147167 & 0.147825 & 0.148395 & 0.148892 \\
			$P_\text{success}$(Qi) &0.144296 & 0.145487 & 0.146363 & 0.147324 & 0.147742 & 0.148371 & 0.148712 \\ 
			$A$ &0.957764& 0.961613& 0.964819 & 0.96753 & 0.969854& 0.971867& 0.973628  \\
			S$(\rho_1)$ & 0.327299& 0.306658& 0.288665 & 0.272835& 0.258792& 0.246247& 0.234962 \\
			S$(\rho_{{\lfloor n/2 \rfloor}})$ &0.705331& 0.701688& 0.706446 &0.703821& 0.707232& 0.705257& 0.707822  \\
			PPT$(\rho_1)$ &-0.237419& -0.22769 & -0.219046 & -0.211304 & -0.204319 & -0.197979 & N/A \\
		\end{tabular}
	\end{ruledtabular}
	\begin{ruledtabular}
		\begin{tabular}{c c c c c c c c c c c c c c}
			n & 17 & 18 & 19 & 20 & 21 & 22 & 23 & 24 & 25 & 26 & 27 & 28 & 29 \\
			\hline\hline
			$P_\text{success}$(Py) & 0.149 & 0.150& 0.150&	0.150& 0.151& 0.151& 0.151 & 0.151 & 0.152 & 0.152 & 0.152 & 0.152 & 0.152 \\
			$P_\text{success}$(Qi) & N/A & N/A & N/A & N/A & N/A & N/A & N/A & N/A & N/A & N/A & N/A & N/A & N/A\\ 
			$A$& 0.975& 0.977& 0.978&	0.979 & 0.980& 0.981& 0.982 & 0.982 & 0.983 & 0.984 & 0.984 & 0.985 & 0.985 \\
			S$(\rho_1)$ & 0.225& 0.215& 0.207&	0.199& 0.192& 0.185& 0.179 & 0.174 & 0.168 & 0.163 & 0.159 & 0.154 & 0.15 \\
			S$(\rho_{{\lfloor n/2 \rfloor}})$ &0.706& 0.708& 0.707&	0.709& 0.708& 0.709  & 0.708 & N/A & N/A & N/A & N/A & N/A & N/A\\
			PPT$(\rho_1)$ & N/A & N/A & N/A & N/A & N/A & N/A & N/A & N/A & N/A & N/A & N/A & N/A & N/A \\
		\end{tabular}
	\end{ruledtabular}
\end{table*}

\end{document}